\documentclass[%
 reprint,%
 amssymb, amsmath,%
 aip,pop,%
]{revtex4-1}

\usepackage{bm}%
\usepackage[colorlinks=true,linkcolor=blue,citecolor=blue,urlcolor=blue]{hyperref}%
\expandafter\ifx\csname package@font\endcsname\relax\else
 \expandafter\expandafter
 \expandafter\usepackage
 \expandafter\expandafter
 \expandafter{\csname package@font\endcsname}%
\fi

\usepackage{graphicx}%

\usepackage{color}
\newcommand{\tr}{\color{red}}

\begin{document}

\def\araa {Annual Review of Astron and Astrophys}
\def\apjl {Astrophysical Journal, Letters}
\def\apj {Astrophysical Journal}
\def\aap {Astronomy and Astrophysics}
\def\ssr {Space Science Reviews}
\def\solphys{Solar Physics}
\def\jgr{J.~Geophys.~Res}

%%%

\title{Resistively-limited current sheet implosions in planar anti-parallel (1D) and null-point containing (2D) magnetic field geometries 
}

%\author[0000-0001-8170-3848]{Jonathan O.~Thurgood}-
%attempting to put the ORCID in doesn't work for me - get bibtex complaints w/ pdflatex
\author{Jonathan O.~Thurgood}
\email{jonathan.thurgood@northumbria.ac.uk}
\affiliation{Division of Mathematics \\
 University of Dundee \\ 
 Dundee 
 \\ DD1 4HN,  UK.}
\affiliation{Department of Mathematics, Physics and Electrical Engineering \\ Northumbria University \\ Newcastle upon Tyne \\ NE1 1ST, UK.}

\author{David I.~Pontin}
\affiliation{Division of Mathematics \\
 University of Dundee \\
 Dundee 
 \\ DD1 4HN,  UK.}

\author{James A.~McLaughlin}
\affiliation{Department of Mathematics, Physics and Electrical Engineering \\ Northumbria University \\ Newcastle upon Tyne \\ NE1 1ST, UK.}

\date{\today}%
\revised{-}%

\begin{abstract}
	Implosive formation of current sheets is a fundamental plasma process. 
	Previous studies focused on the early time evolution, while here our primary aim is to explore the longer-term evolution, which may be critical for determining the efficiency of energy release.
	To address this problem we investigate two closely-related problems, namely: (i) 1D, pinched anti-parallel magnetic fields and (ii) 2D, null point containing fields which are locally imbalanced (`null-collapse' or `X-point collapse'). 
	Within the framework of resistive MHD, we simulate the full nonlinear evolution through three distinct phases: the initial implosion, its eventual halting mechanism, and subsequent evolution post-halting.
	In a parameter study, we find the scaling with resistivity of current sheet properties at the halting time is in good agreement - in both geometries - with that inferred from a known 1D similarity solution.
	We find that the halting of the implosions occurs rapidly after reaching the diffusion scale by sudden Ohmic heating of the dense plasma within the current sheet, which provides a pressure gradient sufficient to oppose further collapse and decelerate the converging flow. 
	This back-pressure grows to exceed that required for force balance and so the post-implosion evolution is characterised by the consequences of the current  sheet `bouncing' outwards. 
	These are: (i) the launching of {propagating} fast MHD waves (shocks) outwards and (ii) the width-wise expansion of the current sheet itself.
	The expansion is only observed to stall in the 2D case, where the pressurisation is relieved by outflow in the reconnection jets.
	In the 2D case, we quantify the maximum amount of current sheet expansion as it scales with resistivity, {and analyse the structure of the reconnection region which forms post-expansion, replete with Petschek-type slow shocks and  fast termination shocks.
	
\textbf{\tr This accepted paper has been published \emph{open access }in Physics of Plasmas, \textbf{25}, 072105 (2018).  It also  contains the animated figures. Please refer to it here: \url{https://doi.org/10.1063/1.5035489}
}	
	} 
\end{abstract}

\maketitle

\section{Introduction} \label{sec:intro}
	Magnetic fields play a key role in determining the dynamics of plasmas at all scales: from fusion experiments and laboratory plasmas to planetary magnetospheres, the Sun and stars, and galaxies and accretion disks. 
	Magnetic reconnection is a fundamental plasma process associated with dynamic energy release in these systems, and it is believed to explain a broad range of phenomena including solar and stellar flares, coronal mass ejections, astrophysical jets, and planetary aurorae. Typically, to be consistent with such systems it is required that the energy release mechanism 
	%must be sudden and explosive.
{	must switch on suddenly (be \lq{explosive}\rq{}).}
	 For rapid  reconnection, we often require the generation of thin layers of intense electric current -- current sheets. As such, the details of how and where such current sheets may be  established are important across a wide range of plasma applications.
	
	One particular mechanism by which current sheets may be established in the vicinity of magnetic null points (or X-type neutral lines with guide fields) is that of \lq{null point collapse}\rq{}, which is an implosive process by which { MHD} waves -- which are generically attracted to null points -- concentrate flux at increasingly small scales producing large current densities. 
	Since the basic idea was first discussed by \citet{dungey53}, the system has been studied extensively using a variety of approaches. 
	Notably, it has been studied dynamically in the close vicinity of the null points (specifically, within the domain close to the null in which the magnetic field and flow can be approximated as linear)\cite{Imshennik1967,syrovatskii1981review,bulanov1984,klapper1996,mellor2003}.
	Such studies tend to indicate unbounded growth of the current in the absence of dissipation.
	Such unbounded growth of current would eventually lead to fast, `explosive' reconnection in any real diffusive system, no matter how small the resistivity. 
	However, since these studies explicitly exclude the surrounding field, it is unclear whether sufficient energy could accumulate at the null in the full system to sustain this current blowup. 
	An alternative approach - to simulate numerically the full nonlinear evolution of the field and flow geometries in response to  a perturbation of fixed total energy in a closed system (i.e., only a finite amount of energy may be supplied to participate in the collapse) - has been considered by a number of authors  \citep[e.g.][]{1991ApJ...371L..41C,1992ApJ...399..159H,1992ApJ...393..385C,1993ApJ...405..207C,1996ApJ...466..487M,2000mare.book.....P,Thurgood2018a}. Such simulations find that collapse is eventually limited either by resistive diffusion, or by the build up of an opposing back-pressure by the associated converging flow, by either plasma compression or compression of an out of plane guide field component.
		
	This approach received significant attention during the 1990s in investigating null collapse as a possible mechanism for obtaining fast reconnection rates in two dimensions, since in the resistively-limited case the scaling of the collapse with decreasing resistivity is suggestive that the implosion can provide fast reconnection and energy release. % \textit{in of itself}.
	These scalings were recently found to mostly extend to the collapse of fully 3D null points  \citep{2017ApJ...844....2T,Thurgood2018a}.
	However,  it was eventually realised that for the solar corona the  ambient plasma pressure is likely sufficient to halt the collapse before the diffusion scale is reached, so that questions were raised over its viability as a fast reconnection mechanism, at least from a solar physics perspective \citep[e.g.][]{1996ApJ...466..487M,2000mare.book.....P}.
	Nonetheless, there are a number of secondary processes which would occur after the initial collapse which could lead to significant energy releases; either secondary current sheet thinning  \citep{1996ApJ...466..487M}, tearing of the current layer depending on its aspect ratio \citep{Thurgood2018a}, or a transition to collisionless reconnection \citep{2007PhPl...14k2905T,2008PhPl...15j2902T,2017JPlPh..83f6301L}. Furthermore, there are other secondary processes that occur after current sheet formation that are of interest -- for example, \textit{Oscillatory Reconnection} \citep{2009A&A...493..227M,2012A&A...548A..98M,2017ApJ...844....2T}, a phenomenon of time-dependent, periodic busts of reconnection, can occur and has been proposed as a possible explanation of quasi-periodic pulsations of solar flares (see also \citet{McLaughlin2018} for a review of the different possible QPP mechanisms)
	As such, the study of null collapse as a means of dynamically forming current sheets in various limiting cases remains of interest, even though the implosion may not itself immediately provide explosive energy release depending on the plasma parameter regime.
	Furthermore,  the behaviour \emph{after} the initial implosion is so far little studied, either analytically (the collapse solutions break down at the singularity), nor computationally (in typical setups used so far, boundary effects and reflections have been communicated to the region of interest by the time at which the initial implosion stalls).
	Investigating such behaviour, isolated from the effects of the boundary, is a key focus of this paper.
	
	A closely-related phenomenon to null collapse is the implosion of planar (1D) current concentrations which are embedded within an anti-parallel magnetic field (i.e. \lq{Harris-like}\rq{} current sheets), which are commonly associated with laboratory pinch experiments. 
	If an electric current is suddenly discharged in an otherwise homogeneous plasma, then there is no pressure gradient available to resist the magnetic pressure gradients within the current sheet and so it implodes in upon itself. 
	Much like the case of null collapse, there exist analytical solutions for cold, ideal plasmas which predict singularity in finite time  \citep[e.g.][]{1982JPlPh..27..491F}.
	 Indeed, in certain limits previous 2D null collapse and 1D collapse solutions have been shown to be equivalent. 
	 For example, the null collapse (2D) similarity solution of \citet{1979JPlPh..21..107F} is obtained analytically by series expansion about  a special case of the initial field perturbation  (their parameter $\epsilon=0$) which allows for a dimensional decoupling in their equations (and hence, an analytic solution) -  which is in fact that reduced to 1D and used in \citet{1982JPlPh..27..491F}. As such, null collapse proceeding from  that initial condition could be regarded as the special limit whereby the imploding, self-similar flow region formed during null collapse along a particular axis becomes identical to that for the self-similar flow region of the 1D pinch  (see also the Appendix of \citet{1982JPlPh..27..491F} for a discussion of the relation of these solutions to those of  \citealt{Imshennik1967}).
	Despite the analytical predictions of singularity in finite time, these implosions are in reality also limited by the eventual progression to small diffusive scales or the formation of back-pressures via the compression, in a direct analogue of the multi-dimensional null collapse case.
	Such limits to the 1D similarity solution were discussed by \citet{1982JPlPh..27..491F} {(and in our Appendix \ref{app:reslimit})}, who also numerically simulated the process in the absence of resistivity, finding good agreement during the initial implosion between the numerics and analytics under those assumptions, although the halting process of the implosion was not properly captured due to insufficient numerical resolution (the outer edge of the current sheet was able to proceed to the grid scale rather than being naturally limited by adiabatic back-pressure). 
	Due to the ease and computational feasibility of placing the outer-boundary sufficiently far from the outer edge of the current sheet in a 1D problem, \citet{1982JPlPh..27..491F} was also able to study the post-implosion behaviour reliably (i.e.~without the interference of boundary reflections) as the rarefaction front which expands outward from the imploding current sheet simply never reached the outer boundary. 
	It was found that immediately after singularity a fast shock front was launched outwards, leaving behind a stationary, thin current sheet. 
	However, with the collapse being halted by a numerical rather than a physical mechanism, it was unclear how physical this behaviour was. 
	 Recently, \citet{2015ApJ...807..159T} revisited the problem, again in the adiabatically-limited case (ideal MHD with finite ambient gas pressure), and confirmed that a shock is launched and a thin current sheet remains in a static state of force balance between the inwardly directed magnetic pressure and outwardly directed gas pressure gradient. Unlike null collapse, the 1D implosion has to our knowledge not been considered in the resistively-limited case which we focus on in this paper.
	 
	The analogy between null collapse and the 1D current sheet implosions stems from a key feature of 2D and 3D null collapse: that during the collapse the field nonlinearly evolves  towards a locally planar, or quasi-1D, geometry. 
	This process has been described by a number of authors \citep[e.g.][]{1979JPlPh..21..107F,syrovatskii1981review,1992ApJ...393..385C} and is the mechanism by which null collapse generates true current sheets (with distinct length- and width-wise axes) even in response to initially cylindrically symmetric current distributions, as we demonstrate later in this paper. 	This process has also been confirmed in laboratory experiments \citep[e.g.][]{1972JETPL..15...94S,2017PlPhR..43..696F}.
	In their numerical study of 2D null collapse, \citet{1996ApJ...466..487M}  realised that the scaling inferred from advancing the analytical 1D similarity solution of \citet{1982JPlPh..27..491F} to the point at which the diffusion and advection terms balance within the induction equation may generally apply to 2D null collapse (further, we note that, since the 1D solution is equivalent to the 2D null collapse solution of \citealt{1979JPlPh..21..107F}  for a special initial condition, that there as at least one instance of null collapse in which we expect this to be true). 
	However, they did \textit{not} find this to be the case, although recently we \citep{Thurgood2018a} noted in a study of 2D and 3D null collapse that our resistively-limited current sheet width scaling appeared to conform closely to 1D scaling of $w\sim{\eta}^{0.89}$ (although we did not compare the scaling of other quantities to the 1D solution).
	The apparent difference between the two sets of simulations is that \citet{1996ApJ...466..487M} allow for resistive diffusion of the flux, but did not then impart that energy upon the plasma via ohmic heating -- suggesting that ohmic heating after reaching the diffusion scale plays a key role in the full halting process. 
	Otherwise, this link is little explored and one we will consider here, for the first time, together with the details of the halting process.

	 In this paper, we study the resistively-limited case of both problems, which is to say setups where the initial, ambient plasma pressure is sufficiently small that it cannot limit the implosions via adiabatic back-pressure before they reach the diffusion scale. 
	Through high-resolution, nonlinear, resistive MHD simulations with effectively open boundaries we are able to simulate the full nonlinear evolution of the 1D and 2D implosions through the three stages of initial implosion, diffusive halting, and the post-halting behaviour (the latter having been considered to date only in the 1D ideal case and not at all in 2D). 
	With such computations we aim to firstly quantify the properties of the current sheets at the time of stalling and test the extent to which scaling inferred from the 1D similarity solution holds for both geometries (given that  the inferred scaling technically only predicts the sheet properties at the time the diffusion scale is reached, which is  only the beginning of the full, nonlinear and diffusive halting process). 
	Second, we wish to examine the precise mechanism by which further implosion is halted after reaching the diffusion scale. 
	Finally we study in detail the properties of the current sheet that remains after the implosion is fully halted-- crucial for understanding how much flux can be reconnected overall as a result of the collapse.
	The paper is structured as follows; first we outline the setup of the simulations (Section \ref{sec:setup}), then detail the behaviour of the initial implosions (Section \ref{sec:implosion}), the mechanism by which the implosion is halted (Section \ref{sec:halting}) and subsequently, the post-implosion evolution (Sections \ref{sec:bounce} and \ref{sec:rec}). Finally we draw conclusions and discuss the results in Section \ref{sec:conclusion}.

%%%%%%%%
\section{Simulation Setup}\label{sec:setup}

\begin{figure*}
\includegraphics[width=18cm]{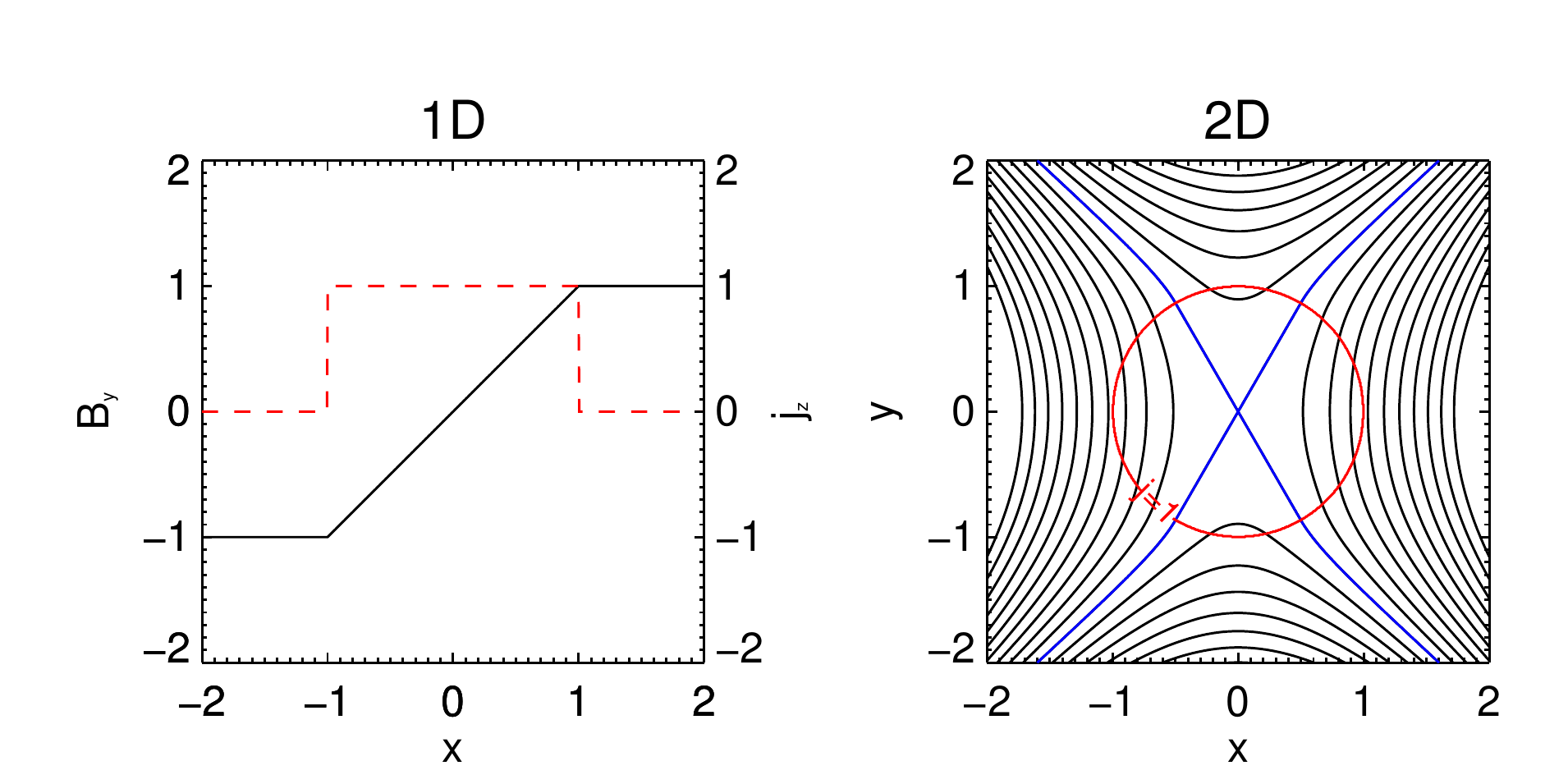} \caption{Initial magnetic fields and current concentrations. Left: The setup of the 1D, planar current sheet implosion indicating the magnitude of the field component $B_y$ (black-solid line) and corresponding initial current density (red-dashed line). Right: The setup of the collapsing null, showing representative fieldlines for the perturbed field where the blue contour indicates the initial location of the separatrix  fieldlines, and red indicates the boundary of the constant, cylindrical current distribution ($j=j_{0}$ inside, and $j=0$ outside). This particular case is illustrated for an exaggerated, larger amplitude perturbation than that used in this paper, in order to make the alteration of the fieldline structure apparent by visual inspection. Both structures are only initially force-free where $r>1$, and so disturbances must be propagated to the boundary at the local fast speed before choices on the boundary affect the evolution. }\label{fig:setup}
\end{figure*}

	The simulations involve the numerical solution of the single-fluid, resistive MHD equations using the LareXd code \citep{2001JCoPh.171..151A}.
	Here we outline the simulation setup (initial conditions), with full technical details deferred to the Appendix. 
	All variables in this paper are nondimensionalized, unless units are explicitly stated, and we set the ratio of specific heats as $\gamma=5/3$ in all simulations.
	
\subsection*{Imploding planar current sheet setup (1D current sheet)}

We use the resistive analogue of the setup of \citet{1982JPlPh..27..491F}  for our 1D simulations. We thus consider the implosion of a pinched current sheet of the form
%
%\begin{eqnarray}
%	B_{y} &=& x \, \mathrm{for} \, |x| \le1 \\
%	&=& \frac{x}{|x|}  \, \mathrm{for} \, x \ge1  \nonumber
%\end{eqnarray}
\begin{equation}
	B_{y} = \left\{
	\begin{array}{ll}
	x~~~~~ & |x| \le1 \\
	\displaystyle\frac{x}{|x|}  & |x| >1,
	\end{array}\right.
\end{equation}
which, as shown in Figure \ref{fig:setup},  corresponds to a uniform plateau of current density out of plane ($j_z$ in our coordinate system) of magnitude $j_0$ = 1 and initial half-width 1. 
	The field is embedded within uniform density plasma ($\rho=1$) at rest ($\mathbf{v}=\mathbf{0}$) with a uniform pressure  $p$ (equivalently,  internal energy density $\varepsilon$) chosen such that the plasma-$\beta$ outside of the current concentration $\beta_e$ is initially low ($\beta_{e}=10^{-8}$). 
	We consider a uniform resistivity $\eta$ as a variable in our study. 
	Under our nondimensionalization, $\eta$ is the value of the inverse Lundquist number as defined by our normalisation constants and so quantifies the relative strength of the diffusivity on the domain-scale.
	 We consider values of $\eta$ in the range $10^{-4}$ to $10^{-2}$, which are sufficiently large relative to the low gas pressure to ensure we consider the resistively-limited regime 
	 { (${\eta}>{3.282}\,{\beta_{e}}^{1.77}$, see appendix \ref{app:reslimit}) }.
	  	As the initial state corresponds to $\mathbf{j}\times\mathbf{B}\neq\mathbf{0}$, and contains no balancing gas pressure gradient, the current concentration immediately implodes. We describe this process in Section \ref{sec:implosion:1}. The external boundary is placed sufficiently far from $x=1$ ({at $x=8$}) that fast waves may not reach it and reflect before {$t=7$}, and so the current sheet evolves as if in a self-consistent, open system  until that time (see Appendix \ref{boundary} for full details).

\subsection*{2D Null Collapse setup}
	We consider the collapse of 2D null points of the Cartesian form 
\begin{equation}\label{eq:2dnull}
\mathbf{B}_{0}=\left[y,x\right]
\end{equation}
which is a potential null point, free from electrical currents, and so constitutes a minimum energy, force-free state. Like in the 1D case, we take the plasma to be initially at rest ($\mathbf{v=0}$), of uniform density ($\rho=1$) and a uniform gas pressure, chosen such that a fixed plasma-$\beta$ defined by the background field $\mathbf{B}_{0}$ at radius $r=1$ may be set, which is taken as $\beta_{0}=10^{-8}$ throughout.  Plasma resistivity $\eta$ is again taken as a uniform variable of our study (in the range $10^{-4}$ to $10^{-2}$).

In order for the 1D and 2D configurations to be as comparable as possible we choose our perturbation to the potential field to give a uniform current out of the plane within a circle of radius 1, with zero current outside. Specifically, we set $\mathbf{B}=\mathbf{B}_{0}+\mathbf{B}^{\prime}$ with
\begin{equation}\label{eq:2dperturbation}
\mathbf{B}^{\prime} =\frac{B_{\theta}\left(r\right)}{r}\left[-y,x\right] ,
\end{equation}
where the component $B_{\theta}$ indicates the radial magnitude of a flux ring
%
%\begin{eqnarray}
%B_{\theta}\left(r\right) &=& \frac{j_{0}r}{2} \, \mathrm{for}\,  r\le1  \\
%&=&  \frac{j_{0}}{2r} \, \mathrm{for}\, r\ge1  \nonumber .
%\end{eqnarray}
\begin{equation}\label{eq:2dperturbation_polar}
	B_{\theta}\left(r\right) = \left\{
	\begin{array}{ll}
 \frac{j_{0}r}{2} \, & r\le1  \\
\frac{j_{0}}{2r} \,& r>1  .
	\end{array}\right.
\end{equation}

	 Thus, on its own, $\mathbf{B}^\prime$ is recognisable as the field associated with a Z-pinch setup. Such cylindrical current concentrations are thought to form in the vicinity of null points due to generic, externally originating MHD waves, which collect near nulls and assume  the cylindrical profile of the Alfv\'en speed isosurfaces due to refraction \citep{2011SSRv..158..205M,2012A&A...545A...9T,2013A&A...558A.127T}.
	 %\citep[see][for a review]{2011SSRv..158..205M}. %verbose version - changed to flow better with PoP  numbered references style
	 Against the background field $\mathbf{B}_{0}$ this perturbation causes the separatrices of the field to be no longer perpendicular, but instead within $r\,{\le}\,1$ assume the angle $\alpha=\mathrm{cos}^{-1}\left(j_{0}/2\right)$. Outside of this region, they asymptotically tend towards their unperturbed positions. 
	This additional flux therefore disrupts force balance within the region $r\le1$ via $\mathbf{j}\times\mathbf{B}\neq\mathbf{0}$; immediately after initialisation MHD waves are launched which establish a system of flow that drives the collapse process via the propagation of this force imbalance. 
	This process is described qualitatively in the following section.
	Again, the outer boundaries are taken to be closed, but are placed sufficiently far away from the vicinity of the initialisation site such that the signal travel time for outward-travelling disturbances which emanate from $r=1$ (the system is force-free at larger radii at $t=0$) to the outer boundary {at $r=20$}  is in excess of the period of interest. This estimated, strict reflection time is $t=\ln{20}\approx2.99$, during which a strictly open, self-consistent evolution is guaranteed, and beyond which it takes further time still for reflected waves to return to the region of interest (typically, $r<1$)  (Appendix \ref{boundary}).
%	Thus, we  model an effectively open system which is arguably the closest null-point containing analogue to that of the 1D planar implosion models, whereby the imploding region and any system of flow established outside of it is established in a self-consistent manner is not effected by reflection of wave fronts, nor any implied (limitless) kinetic or magnetic energy inflow from outside of the domain.

{
We note that presupposing this localized initial disturbance to the magnetic flux is motivated by the well-established result that MHD waves are generically attracted to null points. We thus expect that external perturbations to the larger-scale magnetic fields will preferentially accumulate near nulls, forming current concentrations (see \citet{2011SSRv..158..205M} for a review). Specific examples of this in application can be seen in \citet{2015A&A...577A..70S} and \citet{2017ApJ...837...94T}, where photospheric motions were shown to lead to current accumulation at nulls in realistic model solar atmospheres. In our paper, where we focus on the details of implosive current sheet evolution (i.e., dynamics close to the null rather than those as waves propagate through an  external field and approach the null), we thus presuppose this disturbance as both a matter of computational feasibility and as a modelling simplification, as well as on the preceding physical grounds. This also in line with numerous previous null collapse studies, and allows for closer-comparability to known similarity solutions in 1D and 2D \citep{1979JPlPh..21..107F,1982JPlPh..27..491F}.
}
	
\section{The initial implosion}\label{sec:implosion}

	In this section we describe the qualitative features of the implosions to the time of peak current density (referred to as the critical time or $t_c$). %, which is the point at which the initial implosion and associated increasing  current density is fully halted or stalled. 
	We begin  with a description of the 1D planar implosions, and then consider  nonlinear null collapse. Then, we detail scaling as measured from the 1D and 2D nonlinear simulations for variable $\eta$ in order to determine computationally the extent to which analytically-inferred 1D scaling applies to the full evolution, inclusive of diffusion and the halting process.

\subsection{Imploding planar current sheet} \label{sec:implosion:1}

%%%

\begin{figure*}
\includegraphics[width=18cm]{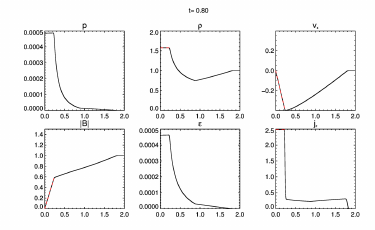}
\caption{ (\textbf{Multimedia view / Animated Figure)} Evolution of fluid and electromagnetic variables along the $x$-axis (horizontal) during the 1D current sheet implosion for the (representative) $\eta=3\times10^{-4}$ case. After initialisation, within the imploding current concentration itself (the plateau in $j_z$) a self-similar evolution is observed until reaching a length scale where diffusion becomes appreciable, which is in excess of $t\approx1$ for all values of $\eta$ considered in this paper. {The similarity region shows excellent agreement with the analytical solution, which is over-plotted with dashed red lines.}
}
\label{fig:1dexample}
\end{figure*}

	The evolution of the simulated 1D implosion before reaching the diffusion scale can be seen in Figure \ref{fig:1dexample} up to $t\approx1.1$ (we delay the discussion of behaviour after this time to later Sections). 
	It is immediately obvious that the initial state (Figure \ref{fig:setup}) is not in force balance, with an unbalanced inwardly directed magnetic pressure gradient. 
	The system immediately responds by launching MHD waves \textit{in both directions}, with the main current concentration collapsing inward, establishing a system of converging mass-flow and flux (this can be considered the `imploding' current sheet).
%	After  the initial time, a system of flow is established. 
	Before the halting process begins in earnest ({ $t \lesssim
1.1$}),  the flow consists of four characteristic regions. 
	Identifying them from left to right (from $x=0$ outward) the first region is the imploding current sheet itself (the plateau in $j_z$) which we identified earlier. It evolves in a self-similar manner, where the decreasing length scale leads to an increasing pinch or gradient, increased current densities, and the converging flow leads to a primarily adiabatic pressurisation (dissipation is negligible until reaching a small enough scale) via the increasing plasma density. 
	It is this region of self-similar flow that is described by the solution of \citet{1982JPlPh..27..491F}, 
	{which is overlaid upon Figure \ref{fig:1dexample} as the red-dashed lines.  We find the analytical and numerical results  to be in excellent agreement until reaching the diffusion scale (where the analytical solution becomes invalid). 
	This solution is considered further in the following sections.}
	Next are two regions which are essentially an expansion or rarefaction as a consequence of the inflowing plasma as driven by the Lorentz force. These regions are separated by a contact surface (most easily identified in the animation of Figure \ref{fig:1dexample} as the minimum in $\rho$) which can be interpreted as the location of the fluid element which initially resides at the edge of the current sheet at $t=0$. Finally, the right-most region is simply undisturbed plasma in its  initial state (the region the rarefaction front is yet to reach), and extends out to the boundary (which is sufficiently far that the system is effectively open for the entire run time).
	Further context regarding these characteristic regions can be gained by comparing Figure \ref{fig:1dexample} to \citet[][Figure 2]{1982JPlPh..27..491F}. The data used for the particular figure is for a run with   $\eta=3\times10^{-4}$, however is generally representative of all values of $\eta$ before the diffusion scale is reached. Before reaching this scale, dissipation is essentially negligible and so the difference in the solutions is minimal. 
	{As such, we achieve the aforementioned agreement with the ideal analytical solution for the similarity region and further, we note agreement with the numerical results in all regions of flow with the ideal simulation of  \citet{1982JPlPh..27..491F} by (visual) comparison of our numerical results at $t=0.8$ to his Figure 3a.
	}

\subsection{Nonlinear, resistively-limited null collapse}

\begin{figure*}
\includegraphics[width=18cm]{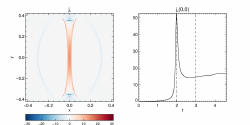}
    \caption{\textbf{(Multimedia View / Animated)} Left: Evolution of current distribution $j_z$ for a 2D collapse proceeding in the nonlinear low-$\beta$ regime ($\eta=3\times10^{-4}$ case).  For sufficiently high amplitudes, the collapse naturally departs from the cylindrical symmetry of the background  wave speed and undergoes quasi-1D evolution due to nonlinearity, forming a true current sheet. 
   % Progression to small-scales still occurs, and so the initial current enhancement  is still focused in to the null as time evolves (note the changing $x$ and $y$ scales). 
    Behaviour after the critical time is also shown, where we see that a fast wave is ejected and that the current sheet undergoes some expansion with a concomitant reduction in current density (which remains enhanced relative to the initial value). 
    %Post critical time behaviour is discussed further in Section \ref{sec:bounce}. 
      Right:  Evolution of $j_z$ at null point in time (black solid curve), where the static horizontal line indicates the critical time $t_c$ and the moving horizontal line indicates the time frame when animated. 
         }
         \label{fig:2DNLevo}
\end{figure*}

\begin{figure*}
\includegraphics[width=18cm]{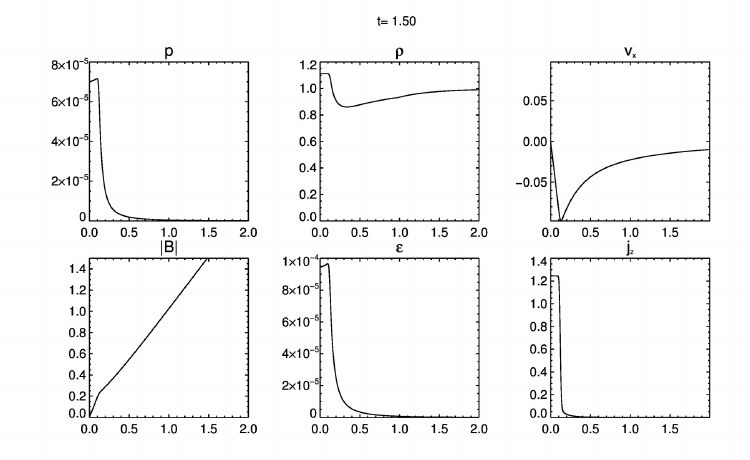}
\caption{ (\textbf{Multimedia View /Animated}) Evolution of fluid and electromagnetic variables along the $x$-axis (horizontal) of the collapsing 2D null. We see that during the initial stages of the implosion that there is a self-similar region of flow, analogous to the 1D case shown in Figure \ref{fig:1dexample}. As in 1D, during the initial phase the data is essentially identical for all $\eta$. During and after the halting, it quantitatively varies with $\eta$, although all cases display the same post-halting behaviour in a qualitative sense (namely, some degree of current sheet expansion, and the ejection of fast waves). The $\eta=3\times10^{-4}$ case is shown. }
	\label{fig:2dxcut}
\end{figure*}

For the 2D configuration in Figure \ref{fig:setup}, the evolution after $t=0$ can be understood in terms of the propagation of the perturbation throughout the domain as MHD waves.
	The excess flux of the perturbation immediately splits into incoming (towards the null) and outgoing (away from the null) characteristics with visible fronts emanating from the boundary of the force-imbalanced and force-free region (i.e., the edge of the initial current distribution), in a similar fashion to the 1D case.
	Due to the arrangement of the Lorentz force, the incoming region establishes a \lq{hyperbolic}\rq{} fluid flow typical of reconnection, with the null itself being a stagnation point separating symmetric and anti-symmetric  regions of inflow and outflow in different regions divided by the separatrices.
	It is the wave-focusing of the incoming  excess  flux  (and the associated current density and Lorentz force-driven flow, both of which will increase in magnitude) which is at the heart of null collapse.

	In low-$\beta$ plasmas, the incoming wave propagates predominantly as  a (magnetically dominated) fast wave (as does the outgoing front, although we largely disregard it in our discussions from this point onwards given the effectively open setup). 
	Initially, the wave propagation is isotropic (moving both across and along fieldlines) and is dictated by the background Alfv\'en speed profile, which is linear in $r$. 
	As such, the wave, its energy, and its associated flows, are propagated inwards -- according to the linearly decreasing wave speed -- and concentrated at increasingly small scales.
	 In the absence of dissipation,{ total current  is conserved and so }the magnitudes of the associated quantities (such as the current density, and the magnitude of Lorentz force, which drives the associated flow) grow and are focused  during this process. 
	 Thus, in this sense, null collapse is a class of MHD implosion with the null being the center of converging magnetic flux, and of plasma compression and rarefaction due to the converging and diverging flow driven by the Lorentz force. 
	 Equivalently, it can be conceptualised as a `Z-pinch' occurring out of plane against a background null-line field, interacting with it.
	As characteristics emanating outside of the null ($r>0$) may not reach  and pass through the null  ($r=0$) at the background Alfv\'en speed ($c_{A}\rightarrow0$ as $r\rightarrow0$),  the implosion continues until some limiting process can grow sufficiently to oppose this focusing. 
	Examples of such processes include resistive dissipation and heating, a growth of plasma ``back-pressure" inside the current concentration due to adiabatic heating, and an analagous magnetic back-pressure due to the presence of a guide field (which, like the plasma itself, is also compressed by the converging flows).  In this paper, we focus on the resistively limited case. 
	
	If the perturbation is sufficiently weak that its flux density does not grow to become comparable to or overwhelm the local background field before reaching the limiting scale (and so begin to evolve nonlinearly), then the entire process is determined by simple advection at the background Alfv\'en speed to the limiting scale.
For this process of \lq{linear null collapse}\rq{}, %During linear null collapse, in the resistively limited case, is characterised by generic perturbations assuming a cylindrical profile due to the Alf\'en speed profile  (i.e. it is not to dependent on the rather symmetric initial conditions we have used) and then maintaining such symmetries until it reaches a scale whereby the diffusion speed exceeds the wave-speed. 
 the associated scaling for current sheet properties at the time of reaching the diffusion scale, has been extensively explored  and we do not consider it further here (see \citealt{2000mare.book.....P} for a 2D overview and \citealt{Thurgood2018a} for 3D extension).
For more energetically significant perturbations
	{ ($|j_{0}|>2\eta$, see Appendix \ref{app:linearamplitude})},
 the increasing perturbation amplitude during the implosion eventually leads the excess flux  carried by the wave to overwhelm the background field and so begin to evolve nonlinearly. 
%	This is likely the case if the collapse is to be energetically relevant in application and that which we consider in this paper.
%	Thus, if perturbations are sufficiently energetic, the collapse eventually enters a  
This nonlinear phase of the collapse is characterised by increasingly planar, quasi-1D behaviour \citep[e.g.][]{1979JPlPh..21..107F,syrovatskii1981review,1993ApJ...405..207C}.
	This is essentially because the perturbation corresponds to regions of total magnetic field enhancement and reduction in  different quadrants either side of the separatrices and so, under a waves interpretation, only certain fronts undergo this nonlinear `acceleration' whilst others begin to stall, providing a means of breaking the initial symmetries in the current distribution. 
	This can also  be understood in terms of the local fieldline structure if one considers the right panel of Figure \ref{fig:setup} (in which $j_{0}$ is taken as an exaggerated, large initial value for visualisation purposes): we can see that the perturbation increases magnetic pressure over magnetic tension in certain quadrants (resulting in an inwardly-directed force), and vice versa in others (resulting in outwardly-directed forces). As the flux {density} (field strength) increases with focusing, this imbalance is enhanced.
	In this manner, nonlinear null collapse produces true current \lq{sheets}\rq{} at null points, as opposed to maintaining the initial cylindrical or ring currents typical of the linear case. 
	The implosion thus becomes increasingly planar, where the nonlinearly accelerated fronts correspond to the converging part of the flow and field, and the stalled  fronts to the length-wise ends of the current layer. %, establishing the outflow jet along current sheet length-wise axis.
 	In the later stages of this quasi-1D evolution, the implosive nature of the collapse is maintained,  becoming  increasingly planar, with a continued focusing and increase in current density (and other quantities) until reaching a scale whereby a limiting process begins, which in this paper is the resistive diffusion scale.
 	
	We can see an example of such evolution in Figure \ref{fig:2DNLevo} (animated), where the initial current density is $j_{0}=0.1$ and $\eta=3\times10^{-4}$. 
	We see that the current sheet soon begins to depart from the initially cylindrical geometry, and becomes increasingly ellipsoidal.
	As the nonlinear acceleration and stalling of the wavefronts in the respective quadrants proceed we can see the process becoming quasi-1D in nature and eventually a rectangular current sheet forms. 
	The sheet continues to thin in this quasi-1D phase until the width is sufficiently small that diffusion becomes appreciable, and so the resistive-halting process begins. We delay discussion of the halting process and post-implosion behaviour, which is visible in the animation, until later sections.
	Furthermore, we can see the evolution of plasma and field variables along the $x$-axis (which becomes the current sheet width-wise axis) before the halting time for the same simulation in Figure \ref{fig:2dxcut}. We see that along this axis, the system of flow and force is qualitatively structured as per the 1D case (compare Figures \ref{fig:1dexample} and \ref{fig:2dxcut}).

\subsection{Scaling at critical time $t_c$}
%{\tb DP: I wonder if we ought to reverse the order of this section and the following one... it's not clear when we are really making these measurements until we see the following bit - alternatively need to explain that better... to discuss??}

\begin{figure*}
\includegraphics[width=18cm]{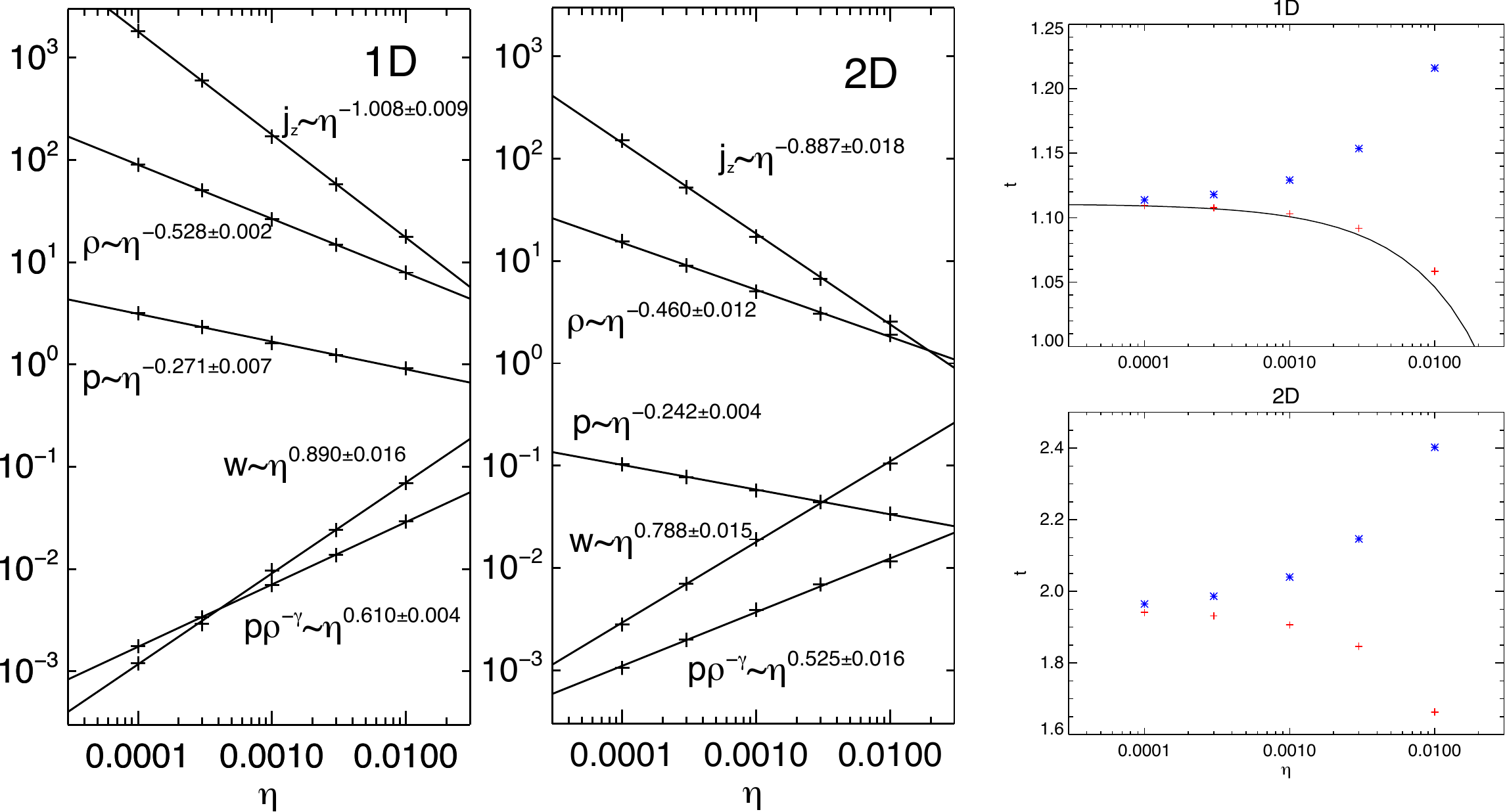}
\caption{
Scaling with resistivity $\eta$ of measured current sheet width $w$, and values at the critical time for $\rho$, $j_z$, $p$ and $p/\rho^{\gamma}$ for the 1D planar implosion at $x=0$ (left panel) and the  2D nonlinear null collapse at the null point (centre panel). The solid lines indicate power-law exponents as determined by a fit to the data points, which are in good agreement with those predicted by the 1D similarity solution.  The $1\sigma$ errors to the fit are shown.  The right panels show the measured critical time $t_{c}$ (blue asterisk) and the time at which resistive halting begins $t_\eta$ (red cross), and the analytically predicted $t_{\eta}$ for the 1D case.}
\label{fig:scaling}
\end{figure*}

	We now consider the current sheet properties at the time at which the implosion stalls {by using the width-wise components of the similarity solution of \citet{1979JPlPh..21..107F} to estimate current sheet properties at the point at which the diffusion region becomes important and the implosion begins to stall.}  
		This reduced, {1D version of the solution as presented in \citet{1982JPlPh..27..491F}}  is as such
\begin{eqnarray}
\sqrt{2}t &=& \rho^{-1} \sqrt{\rho-1} + \mathrm{tan}^{-1} \sqrt{\rho-1} \label{eqn:time} \\
v_{x} &=& -\sqrt{2} \sqrt{\rho-1} \rho x \\ 
B_{y} &=& \rho^{2} x  \\
x_{1} &=& \rho^{-1} \left(\sqrt{\rho}+\sqrt{\rho-1}\right)^{-\sqrt{2}} \, . \label{eqn:width}
\end{eqnarray}
If we require equality of the diffusion and advection terms of the induction equation
\begin{equation}
\frac{\partial}{\partial{x}} (v_{x} B_{y} ) \approx  \eta \frac{\partial^2}{\partial{x^2}} B_{y} 
\label{eqn:inductioncondition}
\end{equation}
then one may determine the value of the outer coordinate $x_1=x_{1}\left(\eta\right)$ (hence, the current sheet half width) at the time at which this condition is met,  and from that all other parameters within the similarity region follow. This may be done either exactly via numerical evaluation, or making use of $\sqrt{\rho-1}\approx\sqrt{\rho}$ (which we have found to be accurate), given we observe a large increase in density before reaching the diffusion scales numerically. Dropping constants, this leads to implied scaling of the current sheet width, $w$, peak current $j$, and mass density, $\rho$, of $w\sim\eta^{0.89}$, $j\sim\eta^{-1.045}$,  $\rho\sim\eta^{-0.5284}$, respectively. 
It is important to note that {strictly speaking} these {scalings} predict the variables at the point in time where the diffusion and advection terms balance
{(which we refer to as the resistive breakdown time $t_\eta$)}.
	However, plasma inertia means that the eventual halting of the collapse would be expected to occur sometime later. 
	This is indeed observed in our simulations. 
	For the purposes of examining scalings of the collapsed current sheet with $\eta$ we define the critical time of the implosion 
	{($t_c$)}	
	to be when the converging inflow within the current sheet is fully decelerated ($v_{x}=0$). This corresponds to the point at which no further current sheet thinning may occur or equivalently where no further growth of current is observed (disregarding secondary effects which may occur post-implosion) and so is also the time where $j_z$ reaches its maximum.

Figure \ref{fig:scaling} shows the scaling of current sheet morphology and the local plasma parameters obtained in our simulations at this time over the range of resistivity $\eta$ considered, for both the planar implosion and the 2D collapse where $j_0=0.1$. 
	We note that this is \textit{not} the special initial condition {for which the solution of \citet{1979JPlPh..21..107F} strictly applies to null collapse} (as mentioned in the introduction) {and so we do not necessarily expect these scalings to apply to the 2D runs}. 
	In our system, that special parameter would be $j_0=2$, a rather extreme case where the initial state corresponding to completely planar field within the region $r<1$. Equivalently, it could be considered to be an initial state where the X-point is fully deformed into a Y-point by the initial condition with zero separatrix angle within $r<1$ (see  Figure 1b of \citealt{1979JPlPh..21..107F} for the field structure in that case). We instead consider a smaller initial current density (where the initial deviation of the  separatrix angle from $\pi/2$ is relatively small) which then evolves naturally towards an increasingly collapsed separatrix angle, with increasingly planar field and increasing current density. We immediately see, from fitting power laws to this data, that both systems obey very similar scaling relationships. 

	In the 1D simulations, it is clear that the analytical prediction is in good agreement with the empirically measured scaling, especially the scaling of $w$ and $\rho$.
	There is a small disagreement in $j$ although we note that $j$ is permitted  to further grow somewhat during the halting process by further thinning and a slight `pile up' of flux from outside of the similarity region at the edge of the current sheet, increasing its magnitude slightly at the edge.
	Aside from this, the fact that the analytical scalings seem to apply at the time of complete stalling -- even though they technically only predict current sheet parameters at the \textit{point at which stalling begins} -- implies that the halting mechanism must be sufficiently rapid to stop the implosion proceeding too much further. This is precisely the observation in our simulations, and is detailed further Section \ref{sec:halting}.
	With regards to the 2D simulations, we find that the measured scaling is similar although not identical. 
	{The reason for the discrepancy in scaling is essentially that our collapses begin with a relatively weak local disturbance to the magnetic field and so undergo a phase of linear, cylindrical null collapse before entering a phase of nonlinear evolution where the current concentration approaches quasi-planarity. This initial phase of null collapse  is not accounted for in the analytically implied scalings (and simply does not occur in the truly planar geometry of the 1D problem}).
	We hypothesise that increasingly energetic null collapses (or equivalently, increasingly weak resistivity $\eta$) will evolve further towards a locally anti-parallel, planar state  before halting begins and so expect that greater perturbations or smaller resistivity will tend increasingly towards the 1D solution, {where if $j_0=2$ the scaling of the two cases becomes identical.} This suggests some utility in applying the 1D solution to null collapse, and we expect that the 1D scaling may be the `upper limit' of scaling (absolute) indices in the resistive regime of  increasingly energetic null perturbations.

%%	{  We find that upon doubling the initial current density to $j_{0}=0.2$, the respective measured scalings are as follows [list],  which is again closer to 1D. Thus, there is some direct numerical evidence in support of this hypothesis.}
	
We have also shown scaling on current sheet pressure $p$ and the quantity $p/\rho^{\gamma}$. The similarity solution uses the  cold plasma approximation and so $p=0$ always (although it does correctly account for the growth of $\rho$ during the initial implosion). We see that the current sheet has become over pressurised by the time of stalling, and that the extent of this pressurisation scales inversely with resistivity, and also that as  $p/\rho^{\gamma}$ is not a constant, that this is not an adiabatic evolution (i.e., it is not an increase in plasma pressure just due to compression). This indicates that Ohmic heating is primarily responsible for this pressurisation. This is also supported by the fact that the simulations of \citet{1996ApJ...466..487M} did report stronger 2D scaling (e.g. $w\sim\eta^{1}$)  in simulations which did not resistively heat the plasma (which is highly compressed in the current sheet), a further clue indicating the importance of ohmic heating for halting the implosion and determining the scale at which that happens altogether. We consider the halting mechanism in detail in the following section.

{
	The right panel of Figure \ref{fig:scaling} shows the measured resistive breakdown times $t_{\eta}$ and critical times $t_{c}$. We note that these times do not obey a power-law but rather asymptote to the (ideal, zero-$\beta$) singularity time as $\eta\rightarrow0$. 
	In the 1D case this time is from equation (\ref{eqn:time}) with $\rho\rightarrow\infty$, yielding $t_{\infty}=\pi / 2\sqrt{2}$ (note this finite-time singularity is referred to as the critical time in other papers, although here we refer to the critical time as the time at which the collapse stalls at a finite scale). Of course, in reality it is expected that as we decrease resistivity some other process would arise to stall the collapse,  such as adiabatic backpressure, at sufficiently small $\eta$. The resistive breakdown time $t_{\eta}$ asymptotically approaches $t_{\infty}$ from below the limit (naturally, as  the consequence of larger diffusion scales with increasing $\eta$) but $t_c$ approaches $t_{\infty}$ from above. This indicates that the \lq{halting time}\rq{} ($t_c - t_\eta$) increases with resistivity.  The 2D case has a larger asymptotic time overall because it initially undergoes an $\eta$-independent  linear phase of evolution before subsequent nonlinear evolution (due to the initially small current density $j_0$). Thus as $j_0$ is increased, the time taken to reach the limiting scales decreases. }

{
	Finally, we note that the 2D reconnection rate ${\eta}{j_z}$ can be inferred from Figure \ref{fig:scaling}, giving a peak reconnection rate of ${\eta}{j_z}\sim{\eta}^{0.113}$. 
	Although this relatively weak scaling could be considered to be suggestive of efficient reconnection, it is only achieved very close to the stalling time (e.g. note the curve of $j_z(t)$ Figure \ref{fig:2DNLevo}). 
	As discussed in the introduction, the average and total reconnected flux during null collapse simulations is not usually found  to scale efficiently with lower $\eta$ (see e.g. Chapter 7 of \citet{2000mare.book.....P}, and Figure 8 of \citet{Thurgood2018a} for total reconnected flux measurements in similar collapse simulations) and so the initial implosion in of itself will not lead to efficient energy release via reconnection in highly conducting plasmas.
	Rather, this may occur by the aforementioned secondary processes occurring after $t_c$.
}

\section{Resistive halting mechanism}\label{sec:halting}

\begin{figure*}
\includegraphics[width=18cm]{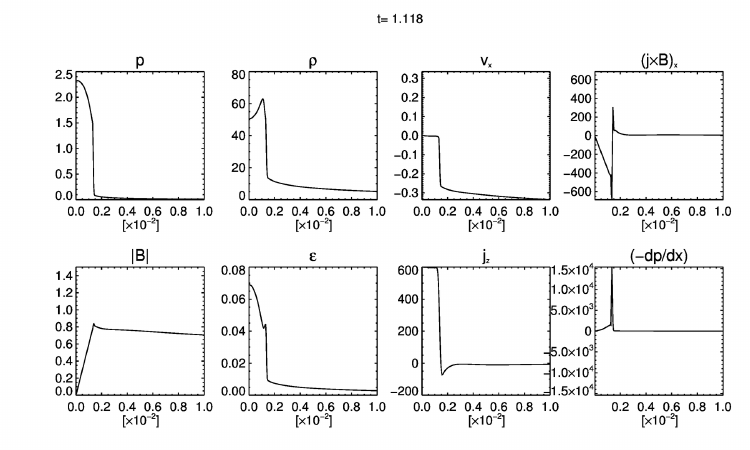}
\caption{ (\textbf{Multimedia View / Animated}) Evolution during the current sheet halting, which occurs rapidly after reaching for the diffusion scale in the $\eta=3\times10^{-4}$, 1D case.  The rapid ohmic heating  of the highly compressed plasma within the current layer provides an internal back-pressure which may oppose the Lorentz force driving the collapse. Once the pressure gradient  matches this force, the inflow begins to decelerate. Further current growth and concomitant ohmic heating continues until the flow is fully decelerated. As such,  the internal pressure is able to overshoot that required for force balance at the time of complete halting. This process is qualitatively similar for variable $\eta$ and in both 1D and 2D geometries. Note that the forces are calculated simply by differencing the pressure and magnetic fields, and so does not give a meaningful  value in the immediate vicinity of a discontinuity. }\label{fig:1dhaltingcuts}
\end{figure*}

\begin{figure*}
\includegraphics[width=18cm]{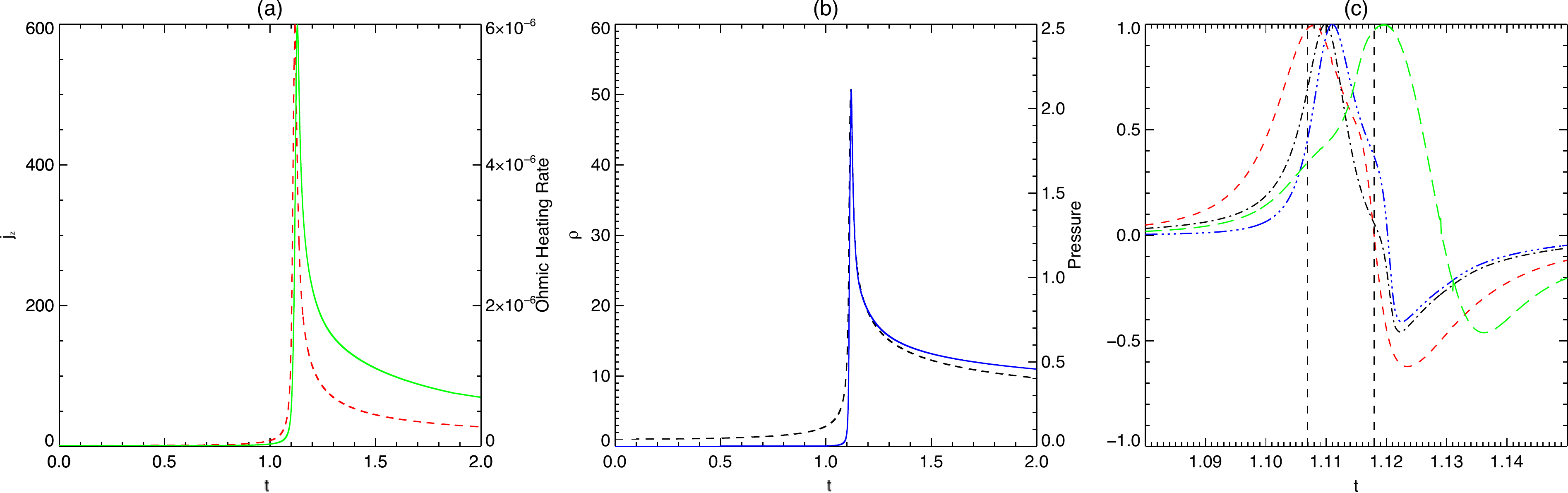}
\caption{Left: Evolution of $j_{z}$ (red-dashed line) at $x=0$ and the instantaneous (simulation-wide) ohmic heating  rate (green-solid line). Centre: Evolution of $\rho(t)$ (black dashed) and $p(t)$ (solid blue) at $x=0$. Both show the highly impulsive nature of the current enhancement, heating, compression  and overall pressurisation once the sheet has proceeded to a sufficiently thin scale. Right: The rate of change of $j_z$ (red linestyle), $\rho$ (black linestyle), $p$ (blue) and ohmic heating (green) (i.e. derivatives of the curves in the left and center figures). The leftmost horizontal line indicates the time at which the analytical solution reaches a scale where diffusion term relevant ($t_{\eta}=1.107$), and the rightmost is the time at which the implosion is completely stopped / sheet thinnest ($t_{c}=1.118$, numerically determined). Thus, $t>1.118$ shows the post-halting evolution of these various quantities.  }\label{fig:1dROChalting}
\end{figure*}

	For both geometries, we have observed that once the imploding current concentration reaches the diffusion scale the process does not immediately stall but, rather, a halting process begins.  
	The evolution of the field and plasma during this process is shown in Figure \ref{fig:1dhaltingcuts} for the 1D, $\eta=3\times10^{-4}$ case, which is qualitatively representative of the halting process in all cases (1D and 2D, all $\eta$ considered). 
	Once the implosion reaches the diffusion scale, there is a loss of similarity within the current sheet as significant ohmic heating begins, increasing the internal energy density and thus plasma pressure within the current sheet dramatically and abruptly, far in excess of what could be achieved by an adiabatic compression alone (note $p/{\rho}^{\gamma}$ is not constant in Figure \ref{fig:scaling}).  
As a result, an outwardly directed pressure gradient develops within the current sheet.
	This quickly becomes comparable in magnitude to, then exceeds, the Lorentz force driving the implosion, and so begins to decelerate the inflowing plasma. 
	During this deceleration process, the pressure gradient continues to grow further still due to both the continued ohmic heating (which even increases in efficacy with continuing growth of $j^2$) and further compression of this now hot, dense current sheet plasma under what remains of the converging flows. 
	By the time of complete deceleration of the converging flow within the current sheet itself ($v_{x}=0$, i.e. the halting time $t_c$), we observe that the interior pressure has grown to be in excess of that required for the establishment of a static current sheet in force-balance -- i.e.~the system overshoots the equilibrium with a force-balanced current sheet. 
	This excess pressurisation means that after the stalling the plasma in the current sheet tries to expand, as discussed in the following section.
	
	The highly abrupt and impulsive nature of the stalling process once the current sheet reaches the sufficiently small scale can be seen clearly in Figure \ref{fig:1dROChalting}, which shows the time-evolution of $j_{z}$, $\rho$, $p$ and the ohmic heating rate at $x=0$, and also shows the time-derivatives of those curves.
	 As small scales are approached, the quantities rapidly rise and the growth rate of current density $j$ only begins to decline immediately after 
	 $t={ {t_{\eta}}}=1.107$ 
	 - the time predicted by advancing the 1D solution \citep{1982JPlPh..27..491F} to the point where diffusion-scale is reached (where Equation \ref{eqn:inductioncondition} is satisfied).  	{We note that until this time $t_\eta$, the growth of variables such as $\rho$ and $j_{z}$ is as predicted by the similarity solution
	 (after this time the similarity solution proceeds to blow-up to singularity, i.e. become invalid), although we exclude these curves from Figure  \ref{fig:1dROChalting} to avoid clutter.}
	There is a short delay until the growth rate of pressure and density begins to decay, the point which we identify with the beginning of the deceleration of the converging flow. 
	 The ohmic heating rate continues to grow throughout the halting process, only beginning to decay in the post-halting evolution (due to current sheet expansion, discussed in the following section). 
	We note that we have found that the qualitative nature of the halting process, as described above for the 1D, $\eta=3\times10^{-4}$ case, is generic to all cases considered in both 1D and 2D.

\section{Post-implosion (bounce)} \label{sec:bounce}

\begin{figure*}
\includegraphics[width=18cm]{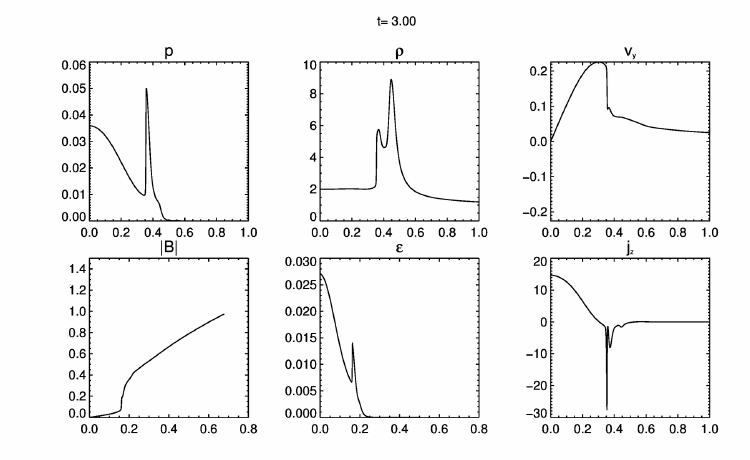}
\caption{ (\textbf{Multimedia View / Animated)} Evolution of fluid and electromagnetic variables along the $y$-axis (horizontal) of the 2D null, for the $\eta=3\times10^{-4}$ case. Post-halting, the internal current sheet pressurisation is relieved by plasma outflow. }
	\label{fig:2dycut}
\end{figure*}

\begin{figure}
\includegraphics[scale=0.4]{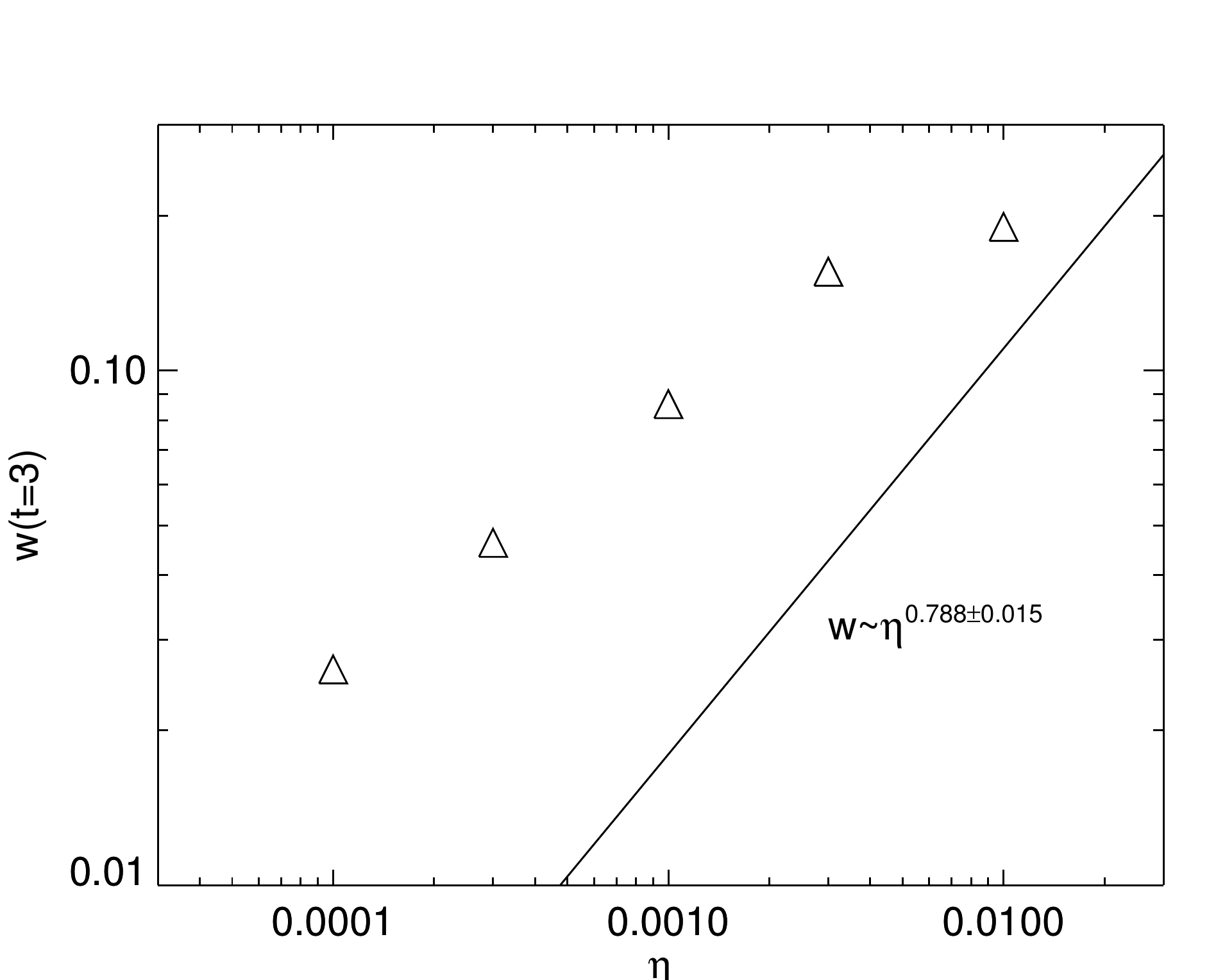}
\caption{ 
Current sheet width  after expansion stalls in the 2D cases (measurements taken at $t=3$). The line plot shows the fit to the width scalings at $t_c$, as per Figure \ref{fig:scaling}.}
	\label{fig:wfit_expansion}
\end{figure}

%%%We have observed that following the resistive deceleration/halting  process (Section \ref{sec:halting}) a current sheet exists which is over-pressurised with respect to the external inwardly-directed forces. 
As a direct consequence of the achievement of an over-pressurised state at the halting time, the current sheet expands outwards into the surrounding plasma in a `bounce'.  This leads to two key features of  post-$t_{c}$ evolution. First is the launching of fast magnetoacoustic waves (or shocks) outward, and second, the width-wise expansion of the current sheet itself. These features are common to all 1D and 2D runs considered, and are visible  in the post-$t_c$ evolution in both  Figure \ref{fig:1dexample} (1D example) and Figure \ref{fig:2DNLevo} (as a 2D example), (both animated). However, comparing 1D and 2D, there are some differences in the evolution of these features.

The first difference between the 1D and 2D implosions is that the outgoing fast waves - which in both cases separate from the current sheet edge at $t>t_c$ and propagate ahead of the expanding current sheet - are only steepened shocks in the 1D case. 
{	
As visible in Figure \ref{fig:1dexample} after $t_c$, in 1D the shock front propagates out into the surrounding plasma, which is  steadily increasing in field strength and density upstream (i.e. increasing upstream magnetic and gas pressures).
 This  shock gradually weakens, and is seen to assume {  a blast-wave like} profile downstream. 
	The launching of the fast shock in 1D is common to the post-implosion evolution simulated under ideal \citep{1982JPlPh..27..491F} {  and} adiabatically-limited cases\cite{2015ApJ...807..159T}. It is also interesting to note that outside the specific context of implosions, quasi-steady current sheets (in force balance with the gas) that become unstable due to the onset of anomalous resistivity also respond with the launching of fast shocks which separate from the current sheet edge, such as that considered by \citet{1982JPlPh..27..157F}. 
}
	 In the 2D case, the outgoing waves 
	 {do  not shock as they separate from the current sheet edge}.
	  We attribute this to the difference in the initial amplitudes $j_{0}$ in the simulations - essentially, the 2D case is less energetic relative to the diffusion, which opposes steeping and shock formation.
	   In 2D, since shocks do not form shortly after $t_{c}$, as {they} propagate out against the {linearly}-growing background Alfv\'en speed,  eventually pulses have a tendency to broaden as their leading edge propagates increasingly ahead of the rest of the waveform 
	{ (based on the linearly increasing background field strength $B_0$, the distance between the leading and trailing edges of a pulse will increase exponentially in time). This, in conjunction with the additional effect of the cylindrical expansion of the pulse, precludes the possibility of steepening and shock formation  once the pulse leaves the immediate vicinity of the diffusion region. We note that the launching of (linear) fast waves from 2D quasi-stable current sheets (again, in response to sudden anomalous diffusion) has been considered in some detail by \citet{2007PhPl...14l2905L}. 
}
	
The second difference {between the 1D and 2D cases }is in the outward expansion of the current sheet boundary itself. 
	This is initially rapid but then slows, though the expansion is only observed to stop in the 2D case (over the time period considered, although we note we simulate the 1D to $t=7$, approximately twice that of the 2D simulations). 
	Given that the 1D expansion must eventually be limited, we attribute the more rapid slowing of the current sheet expansion in 2D to the fact that the over-pressurisation of the current sheet may also be relieved in part by ejection of plasma from the length-wise edges in reconnection jets (which may not form in 1D due to ${\partial}/{\partial}y=0$).
	The development of such a jet for the $\eta=3\times10^{-4}$ example is shown in Figure \ref{fig:2dycut}, which is a cut along the $y$-axis (the sheet's length-wise axis).
	Some outflow is established after initialisation (due to the converging-diverging nature of flow driven directly by the Lorentz force associated with the perturbation); however the key feature is the sudden increase in $\rho$ and $p$ within the current sheet (due to the pressurisation during the implosion) close to the halting time leads to a rapid increase in outflow velocity ($v_y$) inside the sheet, reaching super-magnetosonic speeds (with the jets resembling `wedges' in the 2D plane).
	This results in a substantial amount of mass being ejected out of the current sheet, relieving the internal pressurisation of the current sheet.

The eventual current sheet widths as measured after the expansion stalls at $t=3$ and the fit for thinnest widths from measurements at $t_c$ for the 2D simulations are shown in Figure \ref{fig:wfit_expansion}. The observation is that as resistivity is lowered, current sheets undergo greater relative expansion after the critical time. One factor that influences this expansion is that for smaller $\eta$ a thinner current sheet is obtained at $t=t_{c}$, and therefore during the initial stages of expansion a smaller mass flux is achieved in the outflow jet due to the jet being narrower. Later, as the current sheet expands, it may be that eventually enough plasma can be ejected until the internal pressurisation  is sufficiently relieved. However, a complicating factor is that for different values of $\eta$ we having differing degrees of excess pressurisation in the current sheet at $t=t_c$ (due to changes in the current sheet width, ohmic heating, etc.), recall Figure \ref{fig:scaling}. As such, the expansions for different values of $\eta$  begin from very different starting configurations (at $t=t_{c}$). Therefore the ultimate width-wise expansion of the current sheet is determined by an interplay between different competing factors.	
	
\section{Post-implosion (post-expansion reconnection region)} \label{sec:rec}

\begin{figure*}
\includegraphics[width=18cm]{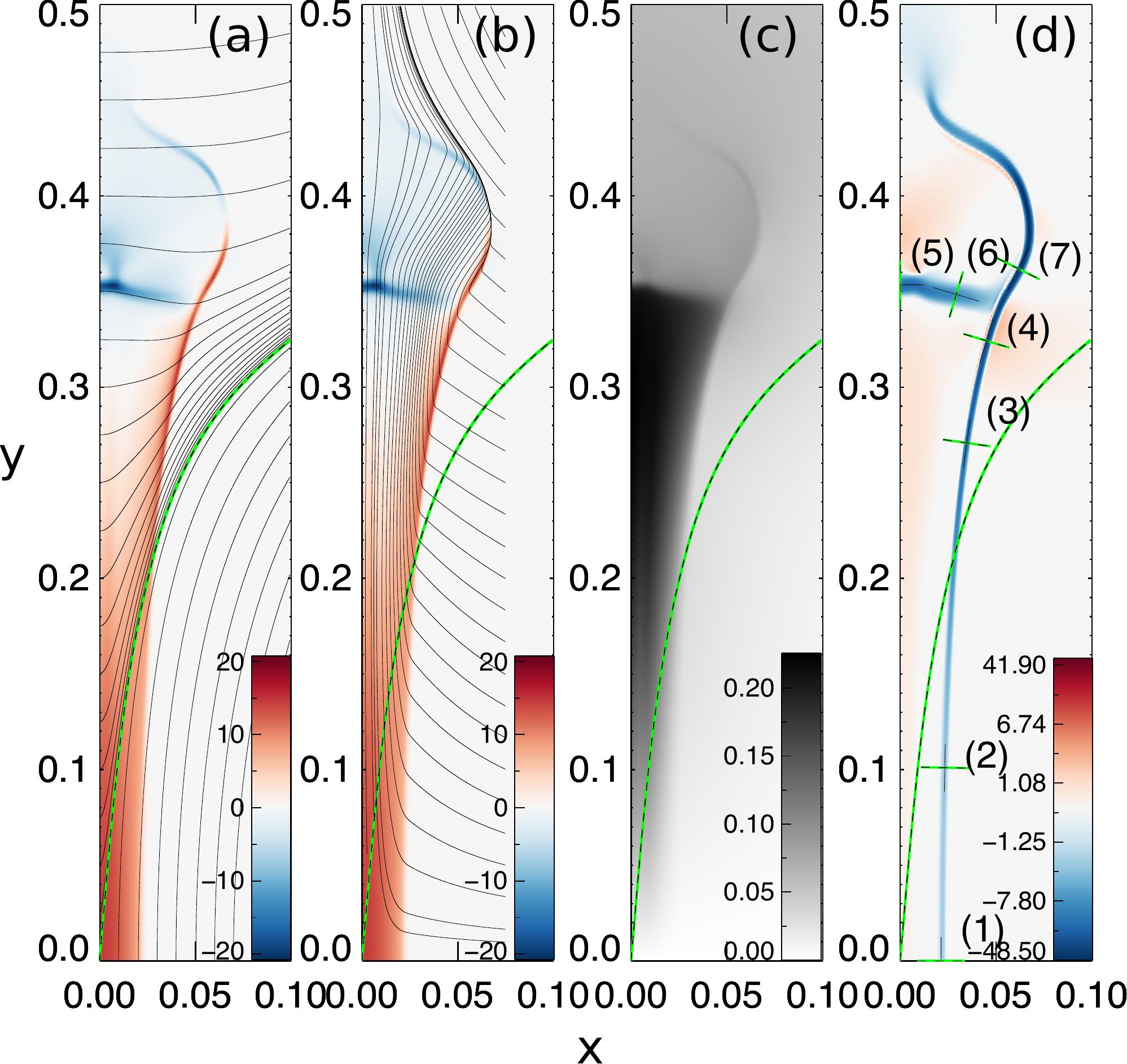}
\caption{ Post-expansion reconnection region in the $\eta=3\times10^{-4}$ case at $t=3$. Against a background of electric current density $j_z$, panel (a) shows magnetic fieldlines and panel (b) shows streamlines as traced from initial points situated along $y=0.075$. Panel (c) $v_{y}$, the dominant component of the outflow jet, illustrating the wedge-shaped profile of the reconnection outflow. Panel (d) shows logarithmically-spaced filled contours of $\nabla\cdot\mathbf{v}$, where the largest calculated compression (deepest blue) outlines the shocks and discontinuities  surrounding the current sheet and expanding jet (red indicating expanding flow), and also the location of the interface normals (and tangents) used to evaluate jumps in Table \ref{table}. 
In all cases, the dashed-black line highlighted by green indicates the separatrix. 
}
	\label{fig:shocksetc}
\end{figure*}

\begin{table*}[]
\centering
\begin{tabular}{ccccccccccccccc}%{lllllllllllllll}
\hline
ID & Type & $x_n$   & $y_n$    & $M_{f1}$ & $M_{f2}$ & $M_{s1}$    & $M_{s2} $   & $M_{A1} $   & $M_{A2}$    & $\rho_{2}/\rho_{1}$ & $B_{2}/B_{1}$ & $p_{t2}/p_{t1}$ & $\theta_1^{\circ}$ & $\theta_2^{\circ}$ \\
\hline
1        & TD         & 1.00 & 0.00  & 0.1 & 0.0 & - & -  & -  & -  & 1.8                & 0.9                          & \textbf{1.0 }                  & 90.0   & 90.0   \\
2        & TD         & 1.00 & -0.02 & 0.1 & 0.1 & 7.3    & 3.2    & 1.3    & 1.1    & 1.7                & 0.9                          & \textbf{1.0}                   & 86.6   & 86.7   \\
3        & SS (P)     & 0.99 & -0.14 & 0.2 & 0.1 & \textbf{8.6}    & \textbf{0.9 }   & 0.7    & 0.4    & 2.7                & 0.9                          & 1.1                    & 71.0   & 69.5   \\
4        & SS (P)     & 0.96 & -0.29 & 0.4 & 0.2 & \textbf{14.6}   & \textbf{0.6 }   & 0.7    & 0.4    & 3.1                & 0.8                          & 1.2                    & 55.1   & 45.9   \\
5        & FS (Perp.) & 0.00 & 1.00  & \textbf{2.0 }& \textbf{0.7 }& - & - & - & - & 2.1                & 2.0                          & 4.1                    & 90.0   & 90.0   \\
6        & FS (Obl.)  & 0.29 & 0.96  & \textbf{1.5} & \textbf{0.8} & 30.6   & 15.0   & 24.4   & 11.5   & 1.5                & 1.5                          & 2.2                    & 84.2   & 83.4   \\
7        & SS (Defl.) & 0.89 & -0.45 & 0.3 & 0.1 & \textbf{18.2 }  & \textbf{0.2}    & 0.4    & 0.1    & 3.7                & 0.9                          & 1.3                    & 43.8   & 34.0   \\
\end{tabular}
\caption{Upstream (subscript 1) and downstream (subscript 2) values measured across  the interfaces indicated in Figure \ref{fig:shocksetc} ($\eta=3\times10^{-4}$ at $t=3$).
% TD corresponds to tangential discontinuities (note the total pressure balance $p_{t2}/p_{t1}$), SS (P) indicates Petschek-like slow shocks, FS indicates the fast termination shock which runs from being perpendicular to oblique, and SS(Defl.) indicates deflected slow shocks past the termination. 
Normal velocities (hence, Mach numbers, with subscript $f$, $s$,and$A$ for those based on fast, slow, and intermediate speeds) are measured in the \textit{laboratory frame}, as the estimated speed of the shocks near this time is sufficiently small to still meaningfully identify transition from super- to sub-magnetosonic flow regions. The symbol $\theta$ indicates the angle  between the magnetic field and the shock normal,  $p_t$ is the total pressure (gas and magnetic), and $(x_{n},y_{n})$ is the normal vector. 
}
\label{table}
\end{table*}
	
	{ % BF for whole section - corresponding right brace at end of this section
			The structure of the { relatively} slowly evolving (but not steady)  reconnection region after the { post-implosion} width-wise expansion stalls is examined   in Figure \ref{fig:shocksetc}, for the specific case of $\eta=3\times10^{-4}$ at $t=3$. 
			It shows magnetic fieldlines and streamlines about the current distribution, the aforementioned super-magnetosonic reconnection jet, and the steep gradients and discontinuities about the jet and diffusion region as {highlighted by} $\nabla\cdot\mathbf{v}$.
%%			(which is calculated by simply differencing $\textbf{v}$ on the grid, and as such does not necessarily give a meaningful number in the vicinity of a discontinuity, but nonetheless outlines its presence). {\tb [Unless you really meant the compressive part???]}
			 A number of { cuts} normal to these interfaces (which are co-spatial with visible features in $j_z$) are indicated, about which jumps in variables across the { interface} are presented in Table  \ref{table}. 
			 We identify four distinct MHD discontinuities present about the reconnection region, namely tangential discontinuities (\lq{TD}\rq{}), standing {slow} shocks which are Petschek-like (\lq{SS(P)}\rq{}), a fast \lq{termination}\rq{} shock (\lq{SF}\rq{}) and a further set of slow shocks (\lq{SS (Defl.)}\rq{}, so-called deflected shocks). 
			 }
		{
			The tangential discontinuity at location [1] (on the $x$-axis) separates the interface between the hot, overdense current sheet plasma and the more rarefied and cool external plasma with total  pressure balance across the transition ($p_{t2}/p_{t1}=1$). 
			Following the edge of the main, nearly-uniform, quasi-1D current concentration  upwards (say, to location [2]), we see similar properties of a steep interface between the current sheet and external plasma with approximate total pressure balance {and close to zero normal field component,} without any super-to-subsonic transition of flow. 
			This balanced interface bounds the main, nearly-uniform, quasi-1D current concentration (we refer to this hereafter as the  \textit{diffusion region}), which extends up to the approximate location at which the separatrix fieldlines intersect with its edge ($y\approx0.2$).}
			
	{	{ Beyond} the end of this diffusion region { ($y\gtrsim 0.2$)}, the wedge-shaped outflow jet - which has now been accelerated to super-magnetosonic speeds - { expands beyond the edge of the diffusion region and impinges upon the inflow. In this region the outflow jet is therefore bounded on its flanks by pairs of shocks. }
		These shocks are identified as slow mode shocks due to the measured jumps at locations [3] and [4], shown in Table  \ref{table}, and the overall effect on fieldlines and streamlines that pass their boundaries visible in Figure \ref{fig:shocksetc}.
		 { Across these shocks the flow transitions from super-to-sub magnetosonic (slow), and the magnetic field is refracted towards the normal, with an overall decrease in field strength. They} are broadly analogous to the 
		 %\lq{switch-off}\rq{}
		  slow shocks in Petschek's model. 
		 The thin, intense current layers which extend outwards from the ends of the main diffusion region, sometimes referred to { as} \lq{bifurcated}\rq{} current sheets,  are manifestations of these shocks, with the currents arising due { to the} field  refraction \cite{1979JPlPh..22....1U,1979JGR....84.7177S}.  		 }

		{In Figure \ref{fig:shocksetc} it is also clear that the reconnection outflow forms a discontinuity at the head of the jet.
		Examination of this shock at locations [5] and [6] indicates that it has the properties of a fast mode shock which runs field-parallel near $y=0$ (tending to a \lq{perpendicular shock}\rq{} as the angle of the field with the normal approaches $\theta=90^{\circ}$) and becomes increasingly oblique as it extends laterally to join the slow shocks on the jets flanks.
		The formation of this shock along the $y$-axis can be observed in the animation of Figure  \ref{fig:2dycut}, where the steepening of $v_y$ into a discontinuity begins at $t\approx2.4$. 
		Despite the rarefied plasma external to the current sheet, the initial outflow (before the shock forms) may not expand unabated as if into a vacuum. 
		Rather, the increasing magnetic and gas pressures will eventually be sufficient to present an obstacle and oppose the pressures driving the jet outward. 
		Thus, as the jet encounters this obstacle, it becomes compressed and a shock forms  (equivalently, the local fast speed increases to the point at which the jet is no longer super-magnetosonic with respect to it, and so with the transition from super-to-submagnetosonic speeds, a shock naturally forms).
		 After its initial formation, the continued effect of this termination shock is to decelerate and heat the continued outflow, with increases in downstream magnetic and gas pressure, as evident at locations [5] and [6] in Table \ref{table}.
		 It exerts a net force against the jet and we observe that the shock slowly propagates inwards towards the null, decreasing the length of the { jet} (visible in the animations of Figures \ref{fig:2DNLevo}  and \ref{fig:2dycut} at later times). 
		 It is likely that no equilibrium position for the termination shock will be reached  due to the unsteady nature of the reconnection inflow.
		  Rather, it is likely that {the longer-term effect} of this inwardly propagating fast shock will be to cause a complete reversal of the current sheet orientation and polarity through a process of \lq{secondary collapse}\rq{} in a manner seen in simulations Oscillatory Reconnection\citep{2017ApJ...844....2T} (specifically, the case of nonlinear, compressive OR where the periodicity is driven by the dynamics local to the diffusion region, as opposed to reflections from a closed boundary as in the classic case of \citet{1991ApJ...371L..41C} and \citet{1992ApJ...399..159H}).		}

{			Finally, we note that downstream of the termination shock { the} region of now-slowly expanding plasma is also flanked by shocks identified as slow shocks (e.g. location [7]). These  slow \lq{deflection shocks}\rq{} are continuations of the Petschek-like shocks past the termination, and are associated with the slow-mode expansion of the ejecta into the surrounding  lower-$\beta$ plasma.
			Following the feature upwards, the sign of the associated current becomes negative due to the change in relative angle between the incident field and the shock normal (towards which field is refracted across a slow shock). At the point at which the angle is zero and the field is entirely normal, the shock is a purely parallel shock and so has no associated current. }

{			The general features described above are common to all values of $\eta$ considered in the post-expansion phase, although we note that typically the slow Petschek shocks appear to be shorter in higher-$\eta$ cases. 
			 It is unclear if this is {primarily} a consequence of the higher reconnection rates with larger $\eta$, or { rather} the fact that $\eta$ { affects the current sheet width as well as the extent to which it is over-dense and over-pressurised,} at the halting time of the implosion (cf Figure \ref{fig:scaling}).
			 { These two effects} cannot be disentangled with the simulations presented in this paper. 
			 Furthermore, a quasi-steady state is not reached and rather at later times still the current sheet shortens along its length-wise axis due to the propagating fast shock discussed previously, and so the length of the Petschek-like slow shocks changes also. These shocks are particularly interesting in that they may allow for further magnetic energy conversion in the post-$t_c$ evolution of the implosively formed reconnection region in a manner reminiscent of the Petschek model, but we defer  investigation of the detailed time-dependent, post-$t_c$ energetics to future studies. 
			   We stress here however that the presence of { Petschek-type} shocks does not in of itself guarantee fast reconnection rates will be achieved \citep{2018arXiv180400324F}. 
			   We also note that, given our uniform resistivity that the very presence of such shocks is most likely due to the time-dependent nature of our problem. Ways in which Petschek reconnection can be achieved in the presence of uniform resistivity have been discussed by a number of authors.\citep{2009PhPl...16a2102B,2009PhPl...16f0701B,2013PhPl...20e2902F,2014PhPl...21k2111B,2018arXiv180400324F}}

\section{Conclusions and Discussion}\label{sec:conclusion}

We have performed numerical simulations of imploding planar current sheets and collapsing magnetic null points in rarefied plasmas of variable resistivity, in order to investigate the full nonlinear evolution of these systems both in the initial stages, up to the stalling, and the subsequent post-implosion behaviour. Here we have addressed the case where the implosion is halted by resistive effects (a future paper will address the implosion as halted by adiabatic back-pressure). Our key findings are as follows:
\begin{enumerate}
\item Empirically, we find that the scaling with resistivity $\eta$  of current sheet parameters produced by the initial implosion are well described by those inferred from the analytical solution to  the diffusion scale, i.e. the point at which the resistive diffusion term of the induction equation grows to equality with the advection  term. 
	The measured scaling for 1D, planar current sheet implosion are nearly identical to the prediction. That this is true, even though the scaling laws technically only predict current sheet parameters at the \textit{point at which stalling begins}, is a direct consequence of the  halting process occurring on a sufficiently short time scale that the implosion may not proceed much further.
	For  2D null point collapses which are sufficiently energetic to evolve nonlinearly  towards a locally quasi-1D or planar implosion we measure scaling that  is slightly weaker than the prediction. 
%	{  Further, we find that the measured scaling  seems to tend towards the 1D scaling as the initial perturbation energy is increased,}
	The weaker scaling is a result of the fact that the initial perturbation  to the null is relatively weak ($j_0$) compared to the  special value required for it to begin in an initially anti-parallel state and so be equivalent to 1D during the self-similar stages of evolution  ($j_{0}=2$). We hypothesise that as the initial perturbation becomes increasingly large that it tends towards the 1D scaling, suggesting they may prove a useful tool for approximating the scaling current sheet properties at $t_c$ for null collapse on the understanding it represents an `upper limit' to the scaling indices in 2D.

%		We find that the initial perturbation for the null point collapse does \textit{not} have to take  the special initial field configuration required for the 2D null collapse of solution \citet[][their parameter $\epsilon=0$]{1979JPlPh..21..107F} to become equivalent to the 1D model used for its application to be a reasonable approximation to the behaviour of the full 2D, nonlinear evolution simulated up to $t_c$. 

\item In both geometries, once the diffusion scale is reached the implosion is halted by the sudden internal pressurisation of the current sheet, which provides a back-pressure to oppose the Lorentz force which drives the implosion inward. 
This pressurisation is a non-adiabatic process (not primarily due to the compression of the current sheet plasma), which is precipitated by the sudden onset of effective Ohmic heating within the current sheet, which is also highly over-dense by this stage due to the compressive nature of the initial implosion.
As high current densities and small scales are maintained during the halting process, the ohmic heating continues throughout and even increases in efficacy. As such, we observe in all cases considered that this heating provides a back-pressure in excess of that needed to simply stall the collapse and achieve force balance. Thus, the system overshoots the force-balanced state.

\item At the instant in which the implosion is fully halted, the current sheet exists in an over-pressurised state. As such, the implosion is immediately followed by a \lq{bounce}\rq{}. This bounce is characterised by both the launching of fast magnetoacoustic waves  outward, and the width-wise expansion of the current sheet, leading to concomitant reduction in current density and associated heating and reconnection rates. We observe the stalling of this width-wise expansion \textit{only in the case of the 2D, null point geometries}, which we attribute to the ability to relieve the over-pressurisation by plasma ejection along the current sheet's length-wise axis,  an effect prohibited in a purely planar geometry. 
\end{enumerate}

{Additionally, we have identified and categorised a number of shock structures which form about the reconnection region in the post-implosion evolution, including Petschek-like slow shocks. These structures may be particularly important for energy conversion occurring after the initial implosion, although we defer a detailed, time-dependent analysis of their energetics to future studies.}

Regarding Key Finding 1, where we find that the 2D scaling is somewhat weaker than that inferred by the 1D similarity solution, we propose that there may be some utility in using  1D approaches (analytical or otherwise) as an approximation to determine current sheet properties at the halting time in more complex field geometries that involve the collapse of null containing structures. 
	Examples may include determining the limiting properties of the merging of colliding magnetic 	islands within tearing current sheets \citep{2015ApJ...807..159T}. 	
	We have considered implosions at 2D nulls  that begin in a state of relatively weak localised collapse in the sense of a small current density or separatrix angle  that evolve naturally towards the planar geometry represented in 1D. These initial values are much smaller perturbations, than that required for the 2D null collapse  solution of \citet{1979JPlPh..21..107F}
to become equivalent to the 1D model examined (we use $j_0=0.1$, as opposed to the special value $j_0=2$). Considering their Figure 1b, it can be readily be appreciated that this special value corresponds to the extreme case of the initial field within $r<1$ beginning in a planar, quasi-1D geometry where within  the current concentration all fieldlines run perfectly parallel. This is stark contrast to say, our Figure 1b ($j_0=1$), which itself is an exaggeration of the values used in this study. 

We note that our proposal that the 1D scalings may represent an upper limit on the (absolute) power of the scalings is somewhat at odds with the low-$\beta$ simulations of \citet{1996ApJ...466..487M}, who considered 2D collapse under similar initial values of current to ours but  reported stronger scaling with resistivity. 
	The most significant difference between their simulations and ours is that while they permit reconnection and field dissipation, they do not self-consistently consider the associated ohmic heating. 
	Rather, their pressure is updated only adiabatically.
	Our interpretation is therefore that their resistive halting mechanism  (which they did not detail) is subtly different to ours.
	In our simulations, as per Key Finding 2, sudden ohmic heating of a highly compressed plasma rapidly stalls the implosion once reaching the diffusion scale.
	We suspect that their implosions may  have proceed somewhat further before being limited directly by dissipation, thus achieving smaller widths at $t_c$ and altering scaling in $\eta$.
	This scaling in $\eta$ is stronger because it therefore prohibits the feedback of the ohmic heating and the plasma compression, thus  increasing the $\eta$ dependence to be closer to their supporting analytic calculations which disregard the plasma density change entirely.
	We also stress that neglecting the ohmic heating is unlikely to affect most other results in that paper (e.g., the finding that higher plasma pressures or guide field strengths can frustrate the possibility of the initial implosion achieving a fast reconnection scaling, but that secondary thinning may occur subsequently).

	The halting as caused by sudden ohmic heating of a compressed current sheet (Key Finding 2) makes for an interesting contrast to the halting of implosions in the purely adiabatic case, which was considered in properly resolved 1D simulations by \citet{2015ApJ...807..159T}.  
	They found that in such a case that the effect of the halting was simply to reflect a shock (which is common to the resistive case here, as per Key Finding 3) and then leave a current sheet behind in a state of force balance. 
	This is very different to the resistive case, where the current sheet expands outwards due to the over-pressurisation, and was only observed to stall in the 2D simulations due to plasma expulsion via the reconnection outflows.
	In some parameter regimes, there may arise the situation where current sheet implosions, whether 1D or 2D, may be limited primarily by an adiabatic process, but still undergo some significant ohmic heating at that halting scale.
	This irreversible magnetic energy dissipation and associated plasma heating introduces an element of inelasticity to the bounce, and so current sheet expansion may be a post-implosion feature even in non-resistively limited regimes.
	We might expect also plasma viscosity to provide for an analogous dissipative halting mechanism, providing for irreversible heating, over-pressurisation and post-implosion similarly to that described in this paper, given it is functionally similar to resistivity in the MHD equations.
	We note that in the solar corona, for instance,  viscosity can be much higher than the resistivity, even of the order $\beta$, and so could play an appreciable role in the halting of such implosions and their post-implosion evolution (see \citep{2010ApJ...725..886C} for a discussion of visco-resistive reconnection). 

	Finally, we note that as in the case of 1D, ideal simulations \citep{1982JPlPh..27..491F,2015ApJ...807..159T}, we find the launching of outwardly directed fast waves in both 1D and 2D). This common feature only apparently differs in that in our 2D setups, the outgoing fast waves do not steepen to shocks, {whereas they do in 1D}.
	We do not think that this is necessarily a generic result, but rather a consequence of the relatively-weaker perturbations we have considered in 2D - in other words, outgoing shocks may form immediately after the implosion in the null collapse geometry  if the implosion is sufficiently energetic. 
%%	
%%	{Although the launching of linear waves from 2D current sheets case has been considered \citep{2007PhPl...14l2905L}, {  to our knowledge there has been no consideration of shocks and their requirements.}
%%}{\tb Not quite sure what you want to say here... Danger of ending on a whimper?}
\acknowledgments

The authors acknowledges generous support from the  Leverhulme Trust and this work was funded by a Leverhulme Trust Research Project Grant: RPG-2015-075. The authors acknowledge IDL support provided by STFC. The computational work for this paper was carried out on HPC facilities provided by the Faculty of Engineering and Environment, Northumbria University, UK. JAM acknowledges STFC for support via ST/L006243/1. DIP acknowledges STFC for support via ST/N000714/1.

\bibliography{references}

\appendix

\section{Nondimensionalisation and the Solver (LareXd code)}\label{appendixA}
{
Following the details  in the LareXd user manual,  the normalisation is through the choice of three basic normalising constants, specifically:
\begin{eqnarray*}
x&=&L_0 \hat{x}\\
\mathbf{B}&=&B_0\hat{\mathbf{B}} \\
\rho&=&\rho_0 \hat{\rho}
\end{eqnarray*}
where quantities with and without a hat symbol are dimensional and nondimensional, respectively. These  are then used to define the normalisation of quantities with derived units through
\begin{eqnarray*}
v_{0}&=&\frac{B_{0}}{\sqrt{\mu_{0}\rho_{0}}}\\
P_{0}&=&\frac{B^{2}_{0}}{\mu_{0}} \\
t_0&=&\frac{L_0}{v_0}\\
j_{0}&=&\frac{B_{0}}{\mu_{0}L_{0}}\\
E_0&=&v_0 B_0\\
%T_0&=&\frac{\epsilon_0 \bar{m}}{k_B}\\
%\mu_{m0}&=&\bar{m}\\
\varepsilon_0&=&v_0^2
\end{eqnarray*}
so that $\mathbf{v}=v_0\hat{\mathbf{v}}$, $\mathbf{j}=j_0\hat{\mathbf{j}}$, $t=t_0\hat{t}$ and 
$P=P_0\hat{P}$ etc. 
% There are further considerations detailed in the manual for temperature normalisations based on average particle mass, but  this is not required in this paper (initial conditions are not prescribed in terms of temperature and we do not calculated temperatures from the simulated data) . 
Applying this normalisation to the ideal MHD equations simply removes the vacuum 
permeability $\mu_0$. In resistive MHD, this scheme leads naturally to a resistivity normalisation:
 \begin{displaymath}
 \hat{\eta}=\frac{\eta}{\mu_0 L_0 v_0}
\end{displaymath}
or $\eta_0=\mu_0 L_0 v_0$. Since $v_0$ is the normalised Alfv\'en speed this means that 
$\hat{\eta}=1/S$ where $S$ is the Lundquist number as defined by the basic normalisation constants. 
}

The simulation is the numerical solution of the nondimensional, resistive MHD equations: ({NB: we drop the hat from this point onwards in the appendix, and throughout the main paper all quantities are nondimensional)}
\begin{eqnarray}
\frac{\mathrm{D}\rho}{\mathrm{D}t}&=&-\rho \nabla\cdot \mathbf{v}\\
\frac{\mathrm{D}\mathbf{v}}{Dt}&=&\frac{1}{\rho}(\nabla\times\mathbf{B})\times\mathbf{B}
-\frac{1}{\rho}\nabla p + \mathbf{F}_{shock}\\
\frac{\mathrm{D}\mathbf{B}}{\mathrm{D}t}&=&(\mathbf{B}\cdot\nabla)\mathbf{v}-\mathbf{B}
(\nabla\cdot\mathbf{v})-\nabla\times(\eta\nabla\times\mathbf{B})\\
\frac{\mathrm{D}\varepsilon}{\mathrm{D}t}&=&-\frac{p}{\rho}\nabla\cdot\mathbf{v}+\frac
{\eta}{\rho}j^{2} + \frac{\mathbf{H}_{visc}}{\rho}\\
\mathbf{j} &=& \mathbf{\nabla}\times\mathbf{B}\\
\mathbf{E} &=& -\mathbf{v}\times\mathbf{B}+\eta\mathbf{j}\\
p &=& \varepsilon\rho\left(\gamma-1\right)
\end{eqnarray}
which are solved on a Cartesian grid using the 2D version of the code (where ${\partial}/{\partial}z=0$ is hard-coded). All results presented are in non-dimensional units. Algorithmically, the code solves the ideal MHD equations explicitly using a Lagrangian remap approach and includes the resistive terms using explicit subcycling \citep{2001JCoPh.171..151A,2016ApJ...817...94A}. The solution is fully nonlinear and captures shocks via an edge-centred artificial viscosity approach \citep{1998JCoPh.144...70C}, where shock viscosity is applied to the momentum equation through $\mathbf{F}_{shock}$ and heats the system through $\mathbf{H}_{visc}$. Extended MHD options available within the code, such as the inclusion of Hall terms, were not used in these simulations. Full details of the code can be found in the original paper \citep{2001JCoPh.171..151A} and the users manual.

\section{Boundary conditions and reflectivity}\label{boundary}

	For the 1D case the equations are solved for the domain $|x|\le8$, although in practice we do not compute a solution in the negative half-space,  but rather exploit appropriate symmetry/antisymmetry conditions on an  `internal'  computational boundary at $x=0$ in order to only calculate for the positive space. At the external boundary ($x=8$) we permit  no flow through or along the boundary ($\mathbf{v}=\mathbf{0}$) with zero-gradient conditions taken on $\rho$ and $\varepsilon$, and also the magnetic field components. In the $y$-direction,  which is 1 cell thick, we set periodic boundary conditions (reducing Lare2D to solve the 1D system with ${\partial}/{\partial}y={\partial}/{\partial}z=0$).
	In 2D,  for the final runs presented here  the simulated domain is the quarter-plane $|x,y|\le20$.
	The boundary conditions on the outer or exterior faces are as in the 1D case - zero-gradient conditions taken $\rho$ and $\varepsilon$, and also on magnetic field components which are tangential to a given face.
	The normal component of the field is held fixed (line-tied) through the boundary.
	Again, appropriate symmetry conditions are exploited along $x=0$ and $y=0$ in order to  only compute on the quarter-plane   $0\le(x,y)\le20$ .
	The suitability of these boundary conditions, and overall stability of the setup, was checked by runs with and without perturbations  (in the null collapse case, recall that the force imbalance is localised to $r<1$). In these tests we found that there was no undesirable behaviour such as the launching of spurious waves from the outer boundary or erroneous current formation at the boundary, and that the state at the boundary remains static until the outwardly propagating part of a given perturbation reaches it. 
	%In the 1D case, as the state at $x>1$ is initially force free, it is obvious by visual inspection that no waves are launched from the outer boundary at $x=8$. 
	The implementation and accuracy of the symmetry conditions were checked simply by re-running some simulations in the whole domain, and we find perfect agreement.

	The outgoing perturbations emanating from the edge of the force-imbalanced region at $x=1$ (1D) or $r=1$ (2D) should not reach the outer boundaries until a fast-speed crossing time. In the 1D case, as the undisturbed Alfv\'en speed is $1$, the reflection time is $t=7$. This is confirmed by inspection of the simulation data - the outgoing front reaches the outer boundary at this time. We take this as the 1D simulation end time and so, the evolution presented is as per system with perfectly open boundary conditions. In the 2D case, we calculate the fast-speed crossing time through the undisturbed medium from $r=1$ to $r=20$ to be  $t=\ln\left({20}/{1}\right)\approx3$, again confirmed by inspection of the data. Thus, our simulations are guaranteed to be consistent with a perfectly open evolution until $t=3$,  in excess of the implosion time and that necessary for expansion to stall. Of course, any reflections must also be propagated back to the region of interest (say, $r<1$) before affecting the dynamics there. It is tempting to suppose the crossing time is doubled on the return, although the propagation is through a disturbed state with regions of inflow, and so the return time could be shorter. Nonetheless, through inspection of the data we are confident that between $t=3$ and the end time of $t=4.7$ that the effect of reflections is minimal, and regardless, all quantitative data presented is measured at times guaranteed to be entirely reflection-free.
 
\section{Grid geometry, resolution and testing}

	In the 1D case, we utilise a uniform grid divided into across the $x$-direction to a maximum resolution of $n_{x}=2\times10^{5}$. 
	The $y$-direction is taken arbitrarily as 1 cell thick and the aforementioned periodic boundary conditions are applied in order to reduce the 2D code to an equivalent 1D code.
	 The sufficiency of the resolution was checked by running many simulations for a given set of parameters with increasing resolution.
	 The agreement presented 1D data and simulations at half the stated resolution is  reassuring, in both  a qualitative sense during the evolution of the implosion and in the sense of producing the same scaling laws (which are in agreement with analytical results, providing for external validation). 	The quantitative difference between quantities measured at $x=0$ is  satisfactory small - in the $\eta=10^{-4}$ case, the most challenging to resolve, the largest quantitative error compared to a simulation at half the resolution is of the order of $4\%$.
	 
	In the case of 2D null collapse, to adequately resolve the small scale features produced by the collapse, especially in the lower resistivity cases, grid stretching is employed to concentrate resolution in the vicinity of the current sheets. The grids cell boundary positions $x_{b}$ along the $x$-direction  are distributed according the transformation \citep{1971LNP.....8..171R,farrashkhalvat2003basic}:
	\begin{equation}
	x_{b} = x_{\mathrm{max}} 
		\frac{\left(\lambda_x+1\right) - \left(\lambda_x-1\right) \left(\frac{\lambda_x+1}{\lambda_x-1} \right)^{1-\xi_x}
		}{\left(\frac{\lambda_x+1}{\lambda_x-1} \right)^{1-\xi_x} +1 }
	\end{equation}
	where $\xi_{x,i}$ is a uniformly distributed computational coordinate $\xi_x\in[0,1]$ subdivided amongst the number of cells used in the $x$ direction. The degree of grid clustering at the $x=0$ is controlled by the stretching parameter $\lambda_x$. Likewise, the same form and parameters are used for the distribution of cells in $y$. 
	In our final 2D simulations of the parameter study, presented here we chose for the $x$-direction $\lambda_x = 1.01$ (more aggressive stretching corresponding to the thin width-wise axis) and $\lambda_y =1.1$ (less aggressive stretching along the length wise axis, but sufficient to ensure cells across the current sheet formed do not possess absurd aspect ratios), then performed simulations with increasing numbers of cells up to a maximum of $nx=ny=4096$,  (effectively, $16384^2$ given the symmetry). 
	Generally, we found that provided the  resolution is sufficient to stop the current sheet collapsing to the grid-scale (i.e., capture the physics of the pressurisation and resistive heating of the current sheet which facilitates the halting process) the solution as measured by the maximal values of current density, density and other variables at the null itself demonstrates convergent behaviour as the numerical resolution is increased. 
	In practice, only the simulations for the smallest resistivity can be ran within a reasonable time at $4096^2$, due to the unfavourable impact of smaller cell sizes upon the resistive timestep (${\Delta}t_{\eta}\propto\Delta_{x}/\eta$). Conveniently, however, higher values of $\eta$ correspond to much wider current sheets at the critical time which therefore do not require such a fine grid (see Figure \ref{fig:scaling}). 
	The final resolution as used for the data presented in 2D the parameter study (Figure \ref{fig:scaling}) is as follows;  ${\eta}=1\times10^{-4}$ uses $n_{x}=n_{y}=4096$  
	yielding ${\Delta}x_{min}\approx0.00013$ and ${\Delta}y_{min}\approx 0.00071$,
	  ${\eta}=3\times10^{-4}$ uses $n_{x}=2048$, $n_{y}=2n_{x}$ yielding ${\Delta}x_{min}\approx0.00026$
	   and ${\Delta}y_{min}\approx 0.00071$, and other values use $n_{x}=1024$ and $n_{y}=2n_{x}$, yielding ${\Delta}x_{min}\approx0.00052$ and ${\Delta}y_{min}\approx 0.00141$. 
	   Each of these final simulations is in good agreement with a simulation at half the stated resolution (half of the cells in each dimension), in a qualitative sense during the evolution of the implosion and in the quantitative sense of producing the same scaling laws (which are also in agreement with analytical predictions and the analytical similarity solution before $t_\eta$). 
	 Like the 1D case, there is an acceptable  level of quantitative agreement with lower resolution simulations, with difference in measured current at the null being of the order of a few percent when compared to the half-resolution case, at worst being the case of  $\eta=10^{-4}$  (the most challenging to resolve) which has the largest difference in measured current density at $t_c$ ($\sim5\%$).

\section{Amplitude required for departure from linear null collapse}\label{app:linearamplitude}
{
The total perturbation energy within the cylinder $r<1$ at $t=0$ is, from the normalised magnetic energy of equation ( \ref{eq:2dperturbation_polar}),
\begin{equation}
	{\delta}E =  \frac{j_{0}^{2}}{8}\int r^2 \mathrm{dA} = \frac{\pi}{16} j_{0}^{2}.
\end{equation} 
This energy is conserved within the similarity region until the solution breaks down (in both linear and nonlinear evolutions), and quickly achieves equipartition thus the total magnetic energy at later times within the imploding region is ${\delta}B={\delta}E/2$. 
The total magnetic energy of the background field $\mathbf{B}_{0}$ (equation \ref{eq:2dnull}) within a cylinder of radius $R$ is calculated as 
\begin{equation}
	U_{B0} (R) = \frac{\pi}{4}R^4 
\end{equation}
Nonlinear evolution will begin to proceed once the perturbation reaches a sufficiently small radius that its own magnetic energy  (and associated magnetic pressure / Lorentz force) becomes comparable to that of the background field. Thus, to enter a nonlinear phase of evolution we require {that $\delta E$ exceeds $U_{B0}$  for some radius $R$ greater} than the linear diffusion radius $r_\eta=\eta^{0.5}$, at which  collapse would otherwise be stalled during its linear evolution ($r_\eta$ is calculated as  the radius at which the background Alfv\'en speed matches the diffusion speed). This yields the condition that $|j_{0}|<2\eta$ for linear, resistively limited null collapse. This condition is broadly consistent with the numerical studies of 2D and 3D null collapse which studied the effect of amplitude on linear versus nonlinear evolution \cite{1996ApJ...466..487M,Thurgood2018a} .	 	 
	}
\section{Conditions for resistively-limited collapse}\label{app:reslimit}
{{The} breakdown of the ideal similarity solution ({ which signals} the beginning of the halting process) { occurs either when the diffusion and advection terms balance}  (in the {  case of resistively limited collapse}, as per {  Equation} \ref{eqn:inductioncondition}) or {  when} the relevant forces balance (for the {  (adiabatic) pressure} limited or  guide field limited cases). Equivalently, one may require comparability of the relative speeds; namely equality of the Alfv\'en speed based on the pinched field $B_{y}$ and the diffusion speed ($\eta/w$), the sound speed ({  with the} assumption of adiabatic evolution to determine $p$), or an {  Alfv{\' e}n} speed associated with the  guide field $B_{z}$ (realising that the guide field within the similarity region will grow with the divergence of flow as $B_{z}=B_{z0}\rho$). Both approaches give the same scaling / power laws.
{  We} obtain the following conditions: 
\begin{eqnarray}
\eta &=& \rho^{3/2}w^2 \\
\beta &=& \left(\frac{2}{\gamma}\right)\rho^{4-\gamma}w^2 \\
B_{z0}  &=& \rho w 
\end{eqnarray}
These may be used to determine the width at the time of breakdown {  of the similarity solution} by introducing equation (\ref{eqn:width}) with the approximation $\sqrt{\rho-1}\approx\sqrt{\rho}$, and then eliminating $\rho$:
\begin{eqnarray}
w_{\eta} &=& 2.156 \eta^{0.892} \\
w_{\beta} &=&  6.222 \beta^{1.579} \\ 
w_{Bz0} &=& 4 B_{z0}^{2.41} 
\end{eqnarray}
where we have taken $\gamma=5/3$.  The dominant limiting process will correspond to whichever of these widths is greatest for a given set of parameters. For resistively limited collapse, we 
{  require $w_\eta>w_\beta\Rightarrow{\eta}>{3.282}\,{\beta}^{1.77}$, and $w_\eta>w_{Bz0}\Rightarrow{\eta}>2{B_{z0}}^{2.707}$.}
}
%find that ${\eta}>{3.282}\,{\beta}^{1.77}$ is required to dominate gas-pressure limiting and ${\eta}>2{B_{z0}}^{2.707}$.

{These conditions should apply to the 1D implosion and 2D collapse as it approaches the extreme of the nonlinear limit (the special initial condition $j_{0}=2$).
% Conditions for the pressure limiting of infinite similarity solutions of \citet{Imshennik1967} may be found in Chapter 7 of  \citet{2000mare.book.....P} (both of which do not apply to the conditions which we simulate in this paper).
}	  
	 
\end{document}